\documentclass[aps,11pt,eqsecnum,amsmath,amssymb,superscriptaddress,showpacs,showkeys,onecolumn,tightenlines]{revtex4}

\usepackage{bm,graphicx,hyperref}
\newcommand{\bra}[1]{\langle\mathord{#1}|}
\newcommand{\ket}[1]{|\mathord{#1}\rangle}
\newcommand{\outerprod}[2]{|\mathord{#1}\rangle\langle\mathord{#2}|}
\newcommand{\innerprod}[2]{\langle\mathord{#1}|\mathord{#2}\rangle}
\newcommand{\matrixel}[3]{\langle\mathord{#1}|\mathord{#2}|\mathord{#3}\rangle}
\newcommand{\proj}[1]{\outerprod{\mathord{#1}}{\mathord{#1}}}
\newcommand{\expect}[2]{\matrixel{#1}{#2}{#1}}

\newcommand{\bbra}[1]{\left\langle\mathord{#1}\right\vert}
\newcommand{\bket}[1]{\left\vert\mathord{#1}\right\rangle}

\newcommand{\Pauli}[1]{\sigma_{#1}}

\begin{document}

\title{Quantum-circuit guide to optical and atomic interferometry}

\author{Carlton M.~Caves} \email{caves@info.phys.unm.edu}
\affiliation{Center for Quantum Information and Control,
University of New Mexico, Albuquerque, New Mexico 87131-0001, USA}

\author{Anil~Shaji} \email{shaji@iisertvm.ac.in}
\affiliation{Indian Institute of Science Education and Research,
Thiruvananthapuram, Kerala 695016, India}

\begin{abstract}
Atomic (qubit) and optical or microwave (modal) phase-estimation
protocols are placed on the same footing in terms of quantum-circuit
diagrams. Circuit equivalences are used to demonstrate the
equivalence of protocols that achieve the Heisenberg limit by
employing entangled superpositions of Fock states, such as N00N
states.  The key equivalences are those that disentangle a circuit so
that phase information is written exclusively on a mode or modes or
on a qubit. The Fock-state-superposition phase-estimation circuits
are converted to use entangled coherent-state superpositions; the
resulting protocols are more amenable to realization in the lab,
particularly in a qubit/cavity setting at microwave frequencies.
\end{abstract}

%06.20.-f: Metrology
%07.60.Ly: Interferometers
%35.25.+k: Atom interferometry techniques
%42.50.St: Nonclassical interferometry, subwavelength lithography
%42.50.Dv: Quantum state engineering
%03.67.-a: Quantum information


\maketitle

\section{Introduction}
\label{sec:intro}

Atomic and optical (or microwave) interferometers are standard tools
for high-precision metrology: the quantity to be measured is mapped
onto the phases of internal levels of atoms or the paths or
polarizations of photons, and the phase is estimated by
interferometry.  The conventional way of operating an interferometer,
using independent (unentangled) atoms or independent (unentangled)
photons, i.e., photons created and processed by the techniques of
linear optics alone, has a sensitivity limit that scales as $1/\sqrt
N$, where $N$ is the number of atoms or photons devoted to the task.
This scaling is called the {\em quantum noise limit\/} or the {\em
shot-noise limit}~\cite{Bachor2004a}.

The sensitivity of an interferometer can be improved by entangling
the atoms or photons.  The ultimate sensitivity is obtained by using
so-called cat states in atomic
interferometry~\cite{Bollinger1996a,Huelga1997a,Shaji2007a} or their
analogue, N00N states, in optical
interferometry~\cite{Gerry2000a,Boto2000a,Gerry2001a,HLee2002a,Gerry2002a,Campos2003a,Gerry2007a,Dowling2008a}.
The moniker N00N, which comes from the (unnormalized) form,
$\ket{N,0}+\ket{0,N}$, was applied and popularized by Dowling and
collaborators~\cite{HLee2002a,Dowling2008a}.  The use of entanglement
permits a sensitivity that scales as $1/N$, a scaling often called
the {\em Heisenberg limit\/}
\cite{Giovannetti2006a,Boixo2007a,Boixo2009au}. There are many
closely related ideas for achieving Heisenberg-limited scaling, some
of which don't look like an interferometer at all (because they
aren't) and don't require any entanglement.  In this paper we use the
pictorial representation of quantum-circuit diagrams to put atomic
and optical interferometry on the same footing and to demonstrate the
equivalence of various methods for achieving the Heisenberg~limit.

In its translations from atoms to photons, this paper formulates a
quantum-circuit version of the equivalence between atomic and optical
interferometry demonstrated by Yurke, McCall, and
Klauder~\cite{Yurke1986a}.  Much of the paper's content on phase
estimation and interferometry is anticipated by the work of Gerry and
collaborators~\cite{Gerry2000a,Gerry2001a,Gerry2002a,Campos2003a,Campos2005a},
and many of the qubit-mode operations discussed in the paper follow a
path blazed by Davidovich, Haroche, and
collaborators~\cite{Brune1992a,Davidovich1993a,Davidovich1996a,Lutterbach1997a,%
Lutterbach1998a,Raimond2001a,Davidovich2004a}.  Lee, Kok, and
Dowling~\cite{HLee2002a} made an initial foray into the domain of
quantum circuits as tools for investigating equivalent atomic and
optical interferometers, calling this the ``quantum Rosetta stone.''
Less lofty, but more ambitious, this article aims to establish a
permanent beachhead on this terrain, making it a standard tool for
analyzing interferometers and phase estimation.

This article is dedicated to the memory of Krzysztof W{\'o}dkiewicz.
When new ideas came along, as in the quantum-information revolution,
Krzysztof embraced them and incorporated them into his suite of
theoretical tools.  Indeed, this contribution is very much in the
spirit of his own recent work on qubit
decoherence~\cite{Wodkiewicz2001a,Daffer2003a,Daffer2004a,Daffer2004b,Dragan2005a},
something he knew as well as anyone, but which he rethought,
re\"examined, and reformulated in the light of the modern language of
quantum operations and Kraus operators.  In the same way, this
contribution repackages concepts from optical and atomic
interferometry, concepts that all physicists know, but presented here
from a different perspective.  It is to be hoped that the paper lives
up to Krzysztof's high standards.  The paper also represents an
experiment that CMC has long wanted to try: writing a paper in which
the figures and captions outweigh the text.

Section~\ref{sec:qubitsmodes} describes atomic interferometry in
terms of qubits and optical interferometry in terms of modes,
reviewing how to map the states and unitary operations for qubits to
those for modes; it introduces the language of quantum circuits, with
particular attention to how measurements are represented and
manipulated within the quantum-circuit picture and the role of
disentanglement in interferometry.  Section~\ref{sec:qnlint} reviews
the description of conventional, quantum-noise-limited
interferometers in terms of circuit diagrams.
Section~\ref{sec:Fockstateint} turns to Heisenberg-limited modal
interferometers.  It considers phase-estimation protocols that are
best thought of as using superpositions of Fock states, and
Sec.~\ref{sec:coherentstateint} discusses the analogous protocols
that use superpositions of coherent states.  A concluding
Sec.~\ref{sec:conclusion} briefly considers how Heisenberg-limited
phase estimation might be implemented in the lab and then wraps up
with a peroration to circuit diagrams as the best way to represent
general protocols for phase estimation and interferometry.

\section{Qubits and modes}
\label{sec:qubitsmodes}

\subsection{States, gates, and quantum circuits}
\label{subsec:statesgates}

Atomic interferometry deals with atoms, idealized as having two
levels.  Any such two-level system can be regarded as a qubit, which
has two standard states, a ``ground'' state $\ket1$ and an
``excited'' state $\ket0$.  Thus we often refer to atomic
interferometry as qubit interferometry.

The pure states of a qubit are conveniently thought of as lying on a
Bloch sphere defined by Pauli operators $Z=\Pauli{z}=\proj0-\proj1$,
$X=\Pauli{x}=\outerprod{0}{1}+\outerprod{1}{0}$, and
$Y=\Pauli{y}=-i(\outerprod{0}{1}-\outerprod{1}{0})$.  We use Latin
letters at the beginning of the alphabet to represent bit values, 0
and 1, and we use Latin letters at the end of the alphabet to denote
the sign representation of a bit value, e.g., $Z\ket a=(-1)^a\ket
a=z\ket a$.  When we have $N$ qubits, we introduce a total ``angular
momentum'' operator $\bm J=\frac{1}{2}\sum_{j=1}^N\bm{\sigma}_j$.  We
denote the product basis of standard states by
$\ket{\bm{a}}=\ket{a_1}\otimes\cdots\otimes\ket{a_N}$, where $\bm{a}$
stands for the bit string $a_1\ldots a_N$.  For qubit interferometry,
we are interested in symmetric qubit states.  The ($N+1$)-dimensional
symmetric subspace is spanned by the states
\begin{equation}
\ket{n_0,n_1}=\ket{\vphantom{\big(}J=N/2,m=(n_0-n_1)/2}=
\sqrt{\frac{n_0!n_1!}{N!}}\sum_{\bm{a}}\ket{\bm{a}}\;,
\end{equation}
where $n_0$ is the number of atoms in state $\ket0$ and $n_1$ is the
number in state $\ket1$, and where the sum runs over all strings $\bm a$
that have $n_0$ 0s and $n_1$ 1s.

In optical or microwave interferometry we deal with two modes of the
electromagnetic field, labeled 0 and 1, with annihilation operators
$a$ and $b$.  Thus we often refer to optical interferometry as modal
interferometry.  The number operators for the two modes are denoted
$N_0=a^\dagger a$ and $N_1=b^\dagger b$.  The Fock basis of
photon-number eigenstates for the two modes is defined by
\begin{equation}
\label{eq:Fockbasis}
\ket{n_0,n_1}_\sim=
\frac{(a^\dagger)^{n_0}}{\sqrt{n_0!}}\frac{(b^\dagger)^{n_1}}{\sqrt{n_1!}}
\ket{0,0}_\sim\;,
\end{equation}
where $\ket{0,0}_\sim$ is the vacuum state of the two modes, and
where $n_0$ is the number of photons in mode~0 and $n_1$ the number
in mode~1.  The connection to qubit interferometry is made by
identifying these Fock states with the symmetric qubit states denoted
in the same way.  The squiggle in Eq.~(\ref{eq:Fockbasis}) is the
symbol we use for modes; we almost always omit it in state
designations.  For $N=1$, the Fock states are the dual-rail logical
states of linear-optical quantum
computing~\cite{Knill2001b,Ralph2002b}, $\ket{1,0}_\sim=\ket0$ and
$\ket{0,1}_\sim=\ket1$.  For $N=2$, the Fock states are
$\ket{2,0}_\sim=\ket{00}$,
$\ket{1,1}_\sim=(\ket{01}+\ket{10})/\sqrt2$, and
$\ket{0,2}_\sim=\ket{11}$.

The identification of symmetric qubits with modes is completed by
introducing the angular-momentum operators of the Schwinger
representation:
\begin{subequations}
\begin{align}
J_z&=\frac{1}{2}(a^\dagger a-b^\dagger b)=\frac{1}{2}\sum_{j=1}^N Z_j\;,\\
J_x&=\frac{1}{2}(a^\dagger b+b^\dagger a)=\frac{1}{2}\sum_{j=1}^N X_j\;,\\
J_y&=-\frac{i}{2}(a^\dagger b-b^\dagger a)=\frac{1}{2}\sum_{j=1}^N Y_j\;.
\end{align}
\end{subequations}
The photon-number operator is $N=a^\dagger a+b^\dagger b$.  The
identification of the $N$-photon subspace of the two modes with the
symmetric subspace of $N$ qubits is the foundation for unifying
optical and atomic interferometry.  Of course, when dealing with
modes, we can also have superpositions of different total photon
numbers.  These superposition states do not have a convenient
representation in terms of qubits, but we will be able to represent
them in our circuit diagrams.

The allowed unitary operations, or gates, are symmetric unitary
operators on the $N$ qubits or, equivalently, photon-number
preserving unitaries for the two modes.  All such unitaries, up to a
global $N$-dependent phase, can be written as $e^{-if(\bm{J})}$,
where $f(\bm{J})$ is a Hermitian function of the total angular
momentum~\cite{fJ}.

An important special case occurs when $f$ is linear, in which case
the unitaries, rotations $e^{-i\bm{J\cdot\hat
n}\theta}=(e^{-i\bm{\sigma\cdot\hat n}\theta/2})^{\otimes N}$ in the
angular-momentum sense, are independent, identical unitaries on the
qubits or, equivalently, the operations of linear optics on the two
modes.  The operation of such a unitary is summarized by how it
transforms the modal creation operators:
\begin{equation}
\label{eq:linopttrans}
e^{-i\bm{J\cdot\hat n}\theta}
\left(\,\begin{matrix}a^\dagger\\b^\dagger\end{matrix}\,\right)\!
e^{i\bm{J\cdot\hat n}\theta}
=M^T\left(\,\begin{matrix}a^\dagger\\b^\dagger\end{matrix}\,\right)\!\;.
\end{equation}
Here $M$ is a $2\times2$ unitary matrix independent of $N$, and $T$
denotes the transpose.   Applying both sides of
Eq.~(\ref{eq:linopttrans}) to the modal vacuum state and converting
to qubit notation gives
\begin{equation}
e^{-i\bm{\sigma\cdot\hat n}\theta/2}
\left(\,\begin{matrix}\ket0\\\ket1\end{matrix}\,\right)\!=
M^T\!\left(\,\begin{matrix}\ket0\\\ket1\end{matrix}\,\right)\!\;,
\end{equation}
from which it follows that $M$ is the representation of
$e^{-i\bm{\sigma\cdot\hat n}\theta/2}$ in the standard qubit basis,
i.e., $M_{ab}=\matrixel{a}{e^{-i\bm{\sigma\cdot\hat n}\theta/2}}{b}$.
The matrix $M$ is given explicitly by
\begin{equation}
\label{eq:M}
M=
\left(\,
\begin{matrix}
\cos(\theta/2)-in_z\sin(\theta/2)&(-in_x-n_y)\sin(\theta/2)\\
(-in_x+n_y)\sin(\theta/2)&\cos(\theta/2)+in_z\sin(\theta/2)
\end{matrix}
\,\right)\!\;.
\end{equation}

\begin{figure*}
\center
\includegraphics[width=6in]{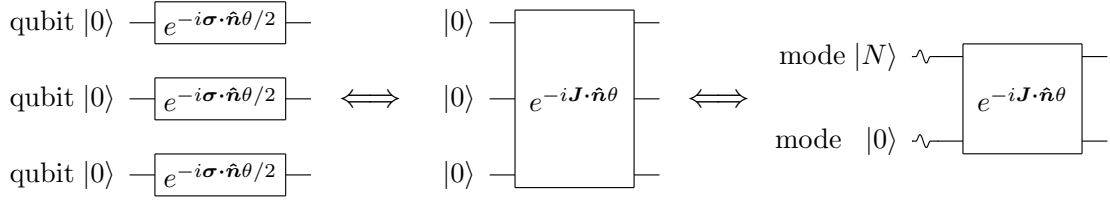}
\caption{Linear-optical quantum gate in qubit form on the left, in
terms of Schwinger operators for qubits in the middle, and in modal
form on the right.  The qubits provide a particle description, and
the modes provide a wave description: the qubits are the
particles---say, atoms or photons---that occupy the modes~0 and 1,
and the modes are the single-particle states, $\ket0$ and $\ket1$, of
the qubits.  In qubit circuits we use $N=3$ qubits for illustration
(occasionally $N=5$), here and in the following, but the equivalences
to modal circuits are valid for arbitrary $N$.  A state to the left
of a circuit wire indicates the input to that wire; if no state is
given, equivalences are independent of the input state, as they would
be here. Equivalences between qubit and modal circuits require,
however, that the input state be in the symmetric subspace of $N$
qubits or, equivalently, in the $N$-photon sector for modes. On the
left and right, wires are labeled explicitly by whether they are
qubits or modes; hereafter, we omit this labeling and use a squiggle,
as in the right circuit, to designate a modal wire.  When $\bm{\hat
n}$ lies in the equatorial plane, these are, for $\theta=\pi/2$,
$\pi/2$ transitions for qubits and 50/50 beamsplitters for modes and,
for $\theta=\pi$, qubit FLIP operators and modal SWAP operators.
\label{fig1}}
\end{figure*}

Suppose $\bm{\hat n}=\bm{\hat x}\cos\phi+\bm{\hat y}\sin\phi$ lies in
the equatorial plane.  If $\theta=\pi/2$, these are~$\pi/2$
transitions for qubits and 50/50 beamsplitters for modes, with
transformation matrix~(\ref{eq:M}) given by
\begin{equation}
M=
\frac{1}{\sqrt2}
\left(\,\begin{matrix}1&-ie^{-i\phi}\\-ie^{i\phi}&1\end{matrix}\,\right)\!\;.
\end{equation}
If $\theta=\pi$, these are qubit FLIP gates and modal SWAP gates,
with additional phase shifts; i.e., $e^{-i\bm{J\cdot\hat
n}\pi}=(-i\bm{\sigma\cdot\hat n})^{\otimes N}$ has transformation
matrix
\begin{equation}
M=
\left(\,\begin{matrix}0&-ie^{-i\phi}\\-ie^{i\phi}&0\end{matrix}\,\right)\!\;.
\end{equation}
Notice also that $Z^{\otimes N}=i^Ne^{-iJ_z\pi}=e^{ib^\dagger
b\pi}=I\otimes\Pi$, with $\Pi$ being the modal parity operator, and
that $(-Z)^{\otimes N}=(-i)^Ne^{-iJ_z\pi}=e^{-ia^\dagger
a\pi}=\Pi\otimes I$.  A $2\pi$ rotation about any axis $\bm{\hat n}$
has $M=I$ and $e^{-i\bm{J\cdot\hat n}2\pi}=(-1)^N=\Pi\otimes\Pi$.

\begin{figure*}
\center
\includegraphics{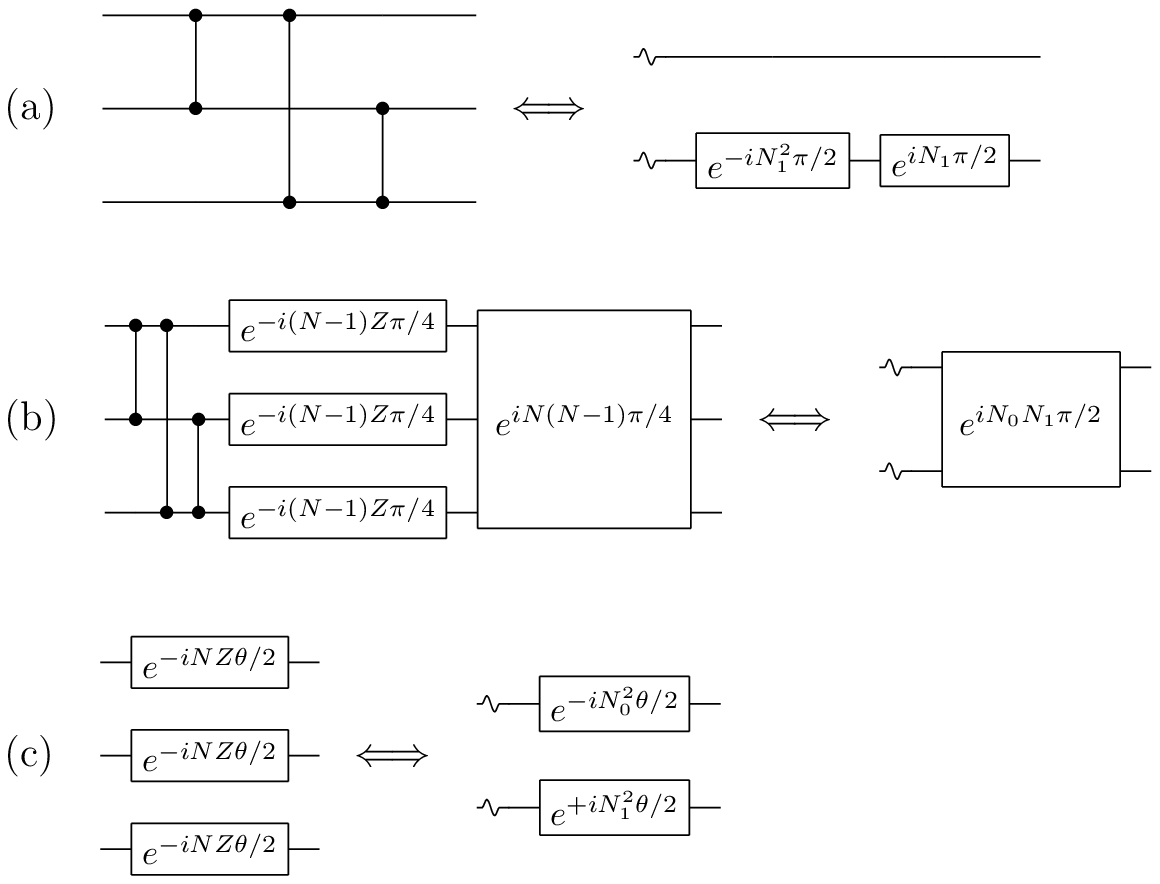}
\caption{Optical Kerr gates and their qubit (particle) equivalents:
(a)~a controlled-SIGN (controlled-$Z$) gate between every pair of
qubits is equivalent to an ordinary $\pi/2$ phase shift on mode~1
plus a $\pi/2$ self-Kerr phase shift on mode~1, i.e., a phase shift
due to a coupling proportional to the square of the number of
photons; (b)~adding $Z$ rotations and an overall phase shift to~(a)
gives a cross-Kerr coupling between modes; (c)~a $Z$-rotation by
angle $N\theta$ on each qubit is equivalent to Kerr phase shifts on
each mode.  As noted in the text, these gates arise naturally in
discussions of nonlinear interferometry based on two-particle
(quadratic) interactions; circuit~(a) appears again in
Fig.~\protect\ref{fig22}(c) as part of a circuit for making
mode-entangled coherent states.  The equivalence in~(a), which is
essentially obvious, can be regarded as coming from the Hamiltonian
identity $H\equiv\frac{1}{4}\sum_{j>k}(Z_j-I)(Z_k-I)=
\frac{1}{2}J_z^2-\frac{1}{2}(N-1)J_z+\frac{1}{4}(N^2/2-N)=
\frac{1}{2}N_1(N_1-1)$; $e^{-iH\pi}$ is a controlled-SIGN gate
between every pair of qubits.  The equivalence in~(b) combines this
Hamiltonian identity with
$\frac{1}{2}N_0N_1=\frac{1}{8}N^2-\frac{1}{2}J_z^2$;~(c) comes from
$\frac{1}{2}(N_0^2-N_1^2)=NJ_z$. Since the gates in these
equivalences are all diagonal in the standard basis, the equivalences
can be mixed without regard to ordering of the gates. \label{fig2}}
\end{figure*}

These considerations provide an opportunity to introduce quantum
circuits in Figs.~\ref{fig1} and~\ref{fig2}. We use use generally
conventional circuit notation~\cite{Nielsen2000a}, with wires
representing qubits or modes and unitary gates enclosed in
rectangular boxes, but we use an initial squiggle to indicate when a
wire applies to a mode.  We make frequent use of the qubit Hadamard
and $S$ gates:
\begin{align}
H&\equiv ie^{-i[(X+Z)/\sqrt2]\pi/2}=\frac{1}{\sqrt2}(X+Z)
\quad\Leftrightarrow\quad
\frac{1}{\sqrt2}
\left(\,\begin{matrix}1&1\\1&-1\end{matrix}\,\right)\!\;,\\
S&\equiv e^{i\pi/4}e^{-iZ\pi/4}
\quad\Leftrightarrow\quad
\left(\,\begin{matrix}1&0\\0&i\end{matrix}\,\right)\!\;.
\end{align}
Notice that $e^{-iY\pi/4}=(I-iY)/\sqrt2=XH=HZ$.  Controlled
application of a unitary is represented conventionally by a circle on
the control qubit connected by a line to the unitary on the target.
Control on $\ket1$ is represented by a filled circle, and control on
$\ket0$ by an open circle.  Figure~\ref{fig1} is the circuit form of
the equivalence between qubit and modal gates for linear-optical
transformations. Figure~\ref{fig2} gives examples of gates and
equivalences outside the linear-optical set; these equivalences lie
at the heart of recent work on nonlinear modal
interferometry~\cite{Boixo2007a,Boixo2009au,Boixo2008a,Woolley2008a,Boixo2008c}.

\begin{figure*}
\center
\includegraphics{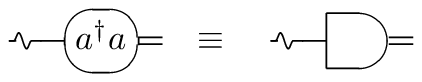}
\caption{Photon counting on a mode, i.e., a measurement of the
photon-number operator $a^\dagger a$, is represented by the
conventional symbol for a photodetector. \label{fig3}}
\end{figure*}

The qubit and modal circuits of Figs.~\ref{fig1} and~\ref{fig2}
illustrate strikingly the difference between entanglement among
qubits, which we call {\em particle entanglement}, and entanglement
between modes, which we call {\em wave entanglement}.  The qubits at
the output of the circuit in Fig.~\ref{fig1} are obviously not
entangled, yet in the case of a 50/50 beamsplitter, for example, the
output modes are entangled: the two modes have anticorrelated,
binomial photon-number distributions in superposition.  A
beamsplitter is an entangling gate for modes, but it is a product
gate for qubits.  In Fig.~\ref{fig2}, the entangling controlled-SIGN
gates in~(a) are equivalent to a product gate for modes; the gates
in~(b), which for qubits differ from those in~(a) only by a set of
local gates, entangle both qubits and modes; the gates in~(c) are
products for qubits and for modes.

We emphasize that qubit and two-mode interferometers can always be
thought of in both ways, as involving particles or modes---that's the
complementarity of quantum mechanics---but we prefer to think one way
or the other, perhaps to our detriment.  We generally like to regard
atomic interferometry as using particles and optical or microwave
interferometry as using modes, but the other way of thinking is
available in both cases.  There is one situation, atomic
interferometry using a two-component Bose-Einstein condensate (BEC),
where it is so natural to think in both ways that it is especially
valuable to be able to translate between the two ways of~thinking.

\subsection{Measurements and circuit equivalences}
\label{subsec:measurements}

\begin{figure*}
\center
\includegraphics[width=6.4in]{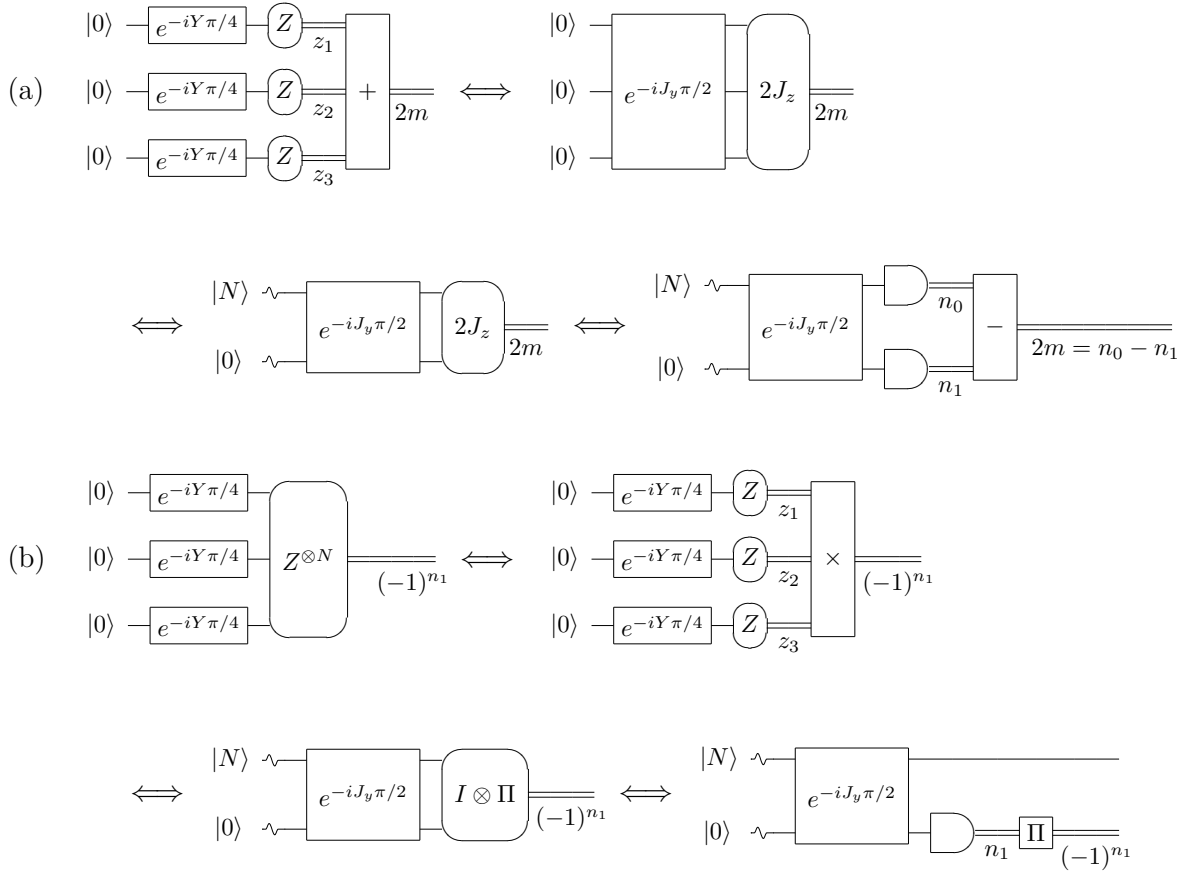}
\caption{(a)~Measurement of $Z$ on $N=3$ qubits, all of which are
prepared in the state $(\ket0+\ket1)/\sqrt2$, followed by classical
processing that outputs the {\em sum\/} of the $N$ outcomes.  The
final outcome $2m=n_0-n_1$ is the difference between the numbers of
qubits found in $\ket0$ and $\ket1$; it is distributed binomially,
with mean zero and variance $N$.  The second and third circuits
transform from qubits to modes via the Schwinger operators, with the
third circuit having $N$ photons in mode~0 processed through a 50/50
beamsplitter and then subjected to a measurement of the
photon-difference operator $2J_z=N_0-N_1$.  In the final circuit, the
differenced photocounting is achieving by photocounting on each mode,
followed by differencing the counts to give outcome $2m=n_0-n_1$.
(b)~Measurement of $Z^{\otimes N}$ on $N=3$ qubits prepared as
in~(a), with outcome $(-1)^{n_1}$ equally likely to be $\pm1$, is
equivalent to measuring $Z$ on all qubits, followed by classical
processing that outputs the {\em product\/} of the $N$ outcomes. This
measurement is equivalent to measuring the parity
$\Pi=(-1)^{b^\dagger b}$ of mode~1.  In the final circuit, the parity
measurement is achieved by photocounting followed by classical
extraction of the parity.  Since mode~0 is unmeasured, it can be
restored as a quantum wire in the final circuit even though this is
not equivalent, strictly speaking, to the third circuit.  The quoted
probabilities for the outcomes depend on the initial state shown, but
the circuit equivalences themselves are independent of initial state.
\label{fig4}}
\end{figure*}

The remaining ingredient needed in our circuit diagrams is a
representation of measurements.  For qubits, we are interested in
symmetric measurements and thus can specialize to measurements of a
particular Pauli component $\bm{\sigma\cdot\hat n}$, usually $Z$, on
all qubits.  We write the $\pm1$ outcome of a measurement of a Pauli
component as $z=(-1)^a$.  Most discussions of qubit measurements
regard the bit value~$a$ as the outcome, but in this paper we always
think of $z$ as the outcome.  For modes, the only measurement we need
is the measurement that counts photons in a~mode.

\begin{figure*}
\center
\includegraphics[width=4.25in]{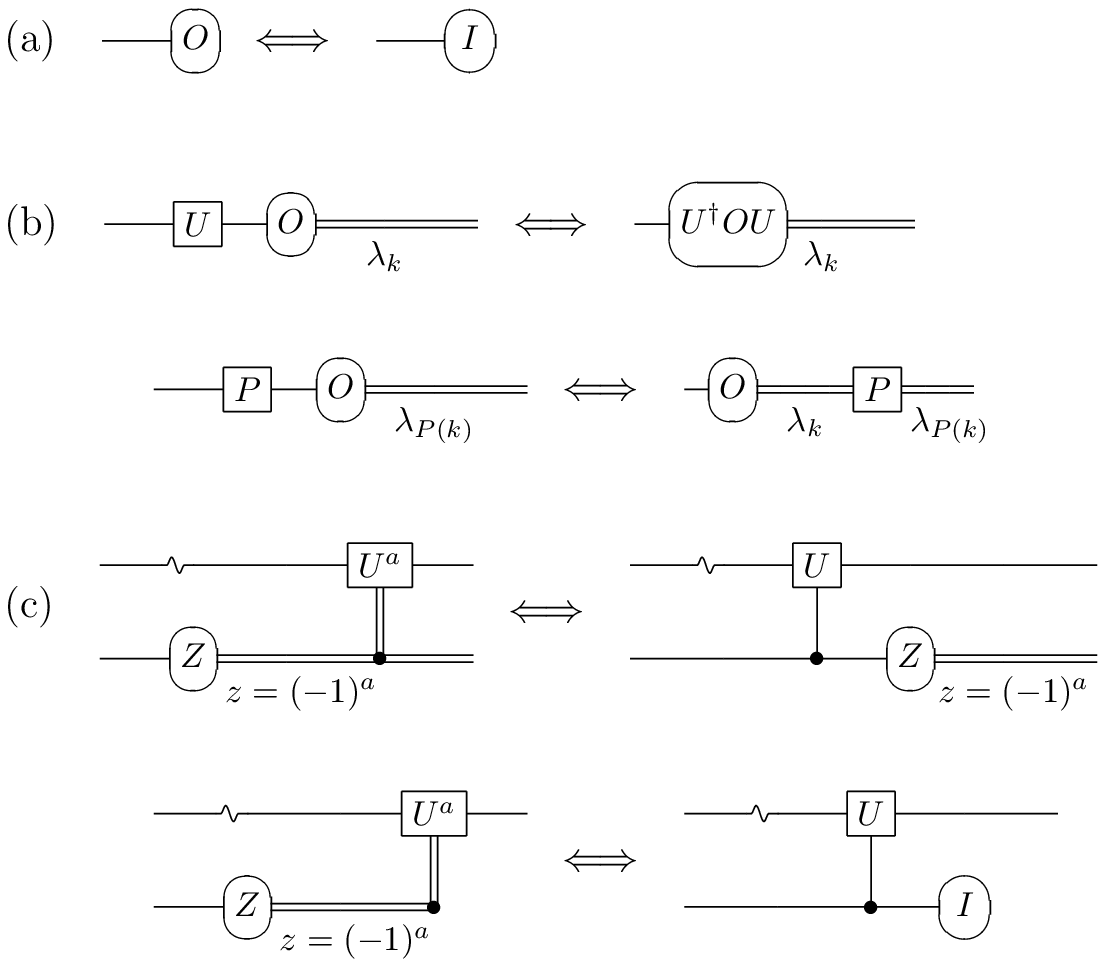}
\caption{(a)~Measurement of an observable $O$, whose outcome is
discarded, is represented by a bubble without a classical wire
emerging from it.  Discarding the outcome implies averaging over the
possible outcomes; this is equivalent to tracing out the measured
system.  We can always represent this as a no-outcome measurement,
i.e., a measurement of the unit operator $I$; since a measurement of
$I$ is really no measurement at all---it merely instructs us to trace
out that system---we can choose to restore the system to the circuit
by eliminating the measurement of $I$ and extending the quantum wire,
{\em provided we remember that the wire must ultimately be terminated
by a measurement without interacting with other wires in the
circuit}. (b)~Top equivalence: Rule for unitarily transforming a
measured observable. Preceding a measurement of observable
$O=\sum_k\lambda_k\ket k\bra k$ with unitary $U$ is equivalent to
measuring the conjugated observable $U^\dagger
OU=\sum_k\lambda_kU^\dagger\ket k\bra k U$; in both cases, incoming
state $\sum_k c_k U^\dagger\ket k$ yields outcome $\lambda_k$ with
probability $|c_k|^2$.  Bottom equivalence: Special case in which $P$
is a unitary that permutes the eigenstates of $O$, i.e., $P\ket
k=\ket{P(k)}$, where $P(k)$ is a permutation of the indices.  In this
situation, $P^\dagger
OP=\sum_k\lambda_k\bket{P^{-1}(k)}\bbra{P^{-1}(k)}
=\sum_k\lambda_{P(k)}\ket k\bra k$; thus a measurement of $P^\dagger
OP$ is the same as measuring $O$ and then applying the permutation to
the outcome. On both sides of the equivalence, incoming state $\sum_k
c_k\ket k$ yields outcome $\lambda_{P(k)}$ with probability
$|c_k|^2$.  (c)~Top equivalence: Principle of deferred measurement,
depicted here for a mode and a qubit. Measuring $Z$ on the qubit and
using the outcome to control classically the application of a unitary
$U$ to the mode is the same as applying a controlled-$U$, with the
qubit as control and the mode as target, and then measuring $Z$ on
the qubit. In the latter case, the control is done coherently, and
the outcome of the measurement later reveals whether the unitary was
applied. Bottom equivalence: Same principle, but with the outcome of
the $Z$ measurement discarded after application of the classical
control. \label{fig5}}
\end{figure*}

In our circuit diagrams, measurements are represented by bubbles
(rounded boxes), with the measured observable inside the bubble.  For
photon counting, we adopt a special notation, illustrated in
Fig.~\ref{fig3}, in which the measurement is represented by the
conventional symbol for a photodetector.  We are only interested in
the outcome of a measurement and its statistics and not in the
post-measurement state of the measured system; in the lingo of
quantum measurement theory, this means we are interested in the
measurement's POVM, but not in the quantum operation that outputs a
post-measurement state.  Thus, in our circuits, there is no quantum
wire emerging from a measurement bubble; the only output from a
bubble is a double (classical) wire that carries the value of the
outcome.  The values on such classical wires can be manipulated by
classical gates, which we depict as rectangular boxes. We often label
a classical wire with the value it carries.

Figure~\ref{fig4} illustrates our measurement notation in the context
of two ways of processing the data from $Z$ measurements on all
qubits and the equivalent measurements on modes.  The important
conclusions are that summing the outcomes of $Z$ measurements is
equivalent to counting photons on the two modes and then differencing
the counts, whereas multiplying the outcomes of $Z$ measurements,
i.e., measuring $Z^{\otimes N}$, is equivalent to measuring the
parity of mode~1~\cite{Gerry2000a,Gerry2001a,Gerry2002a}.

\begin{figure*}
\center
\includegraphics{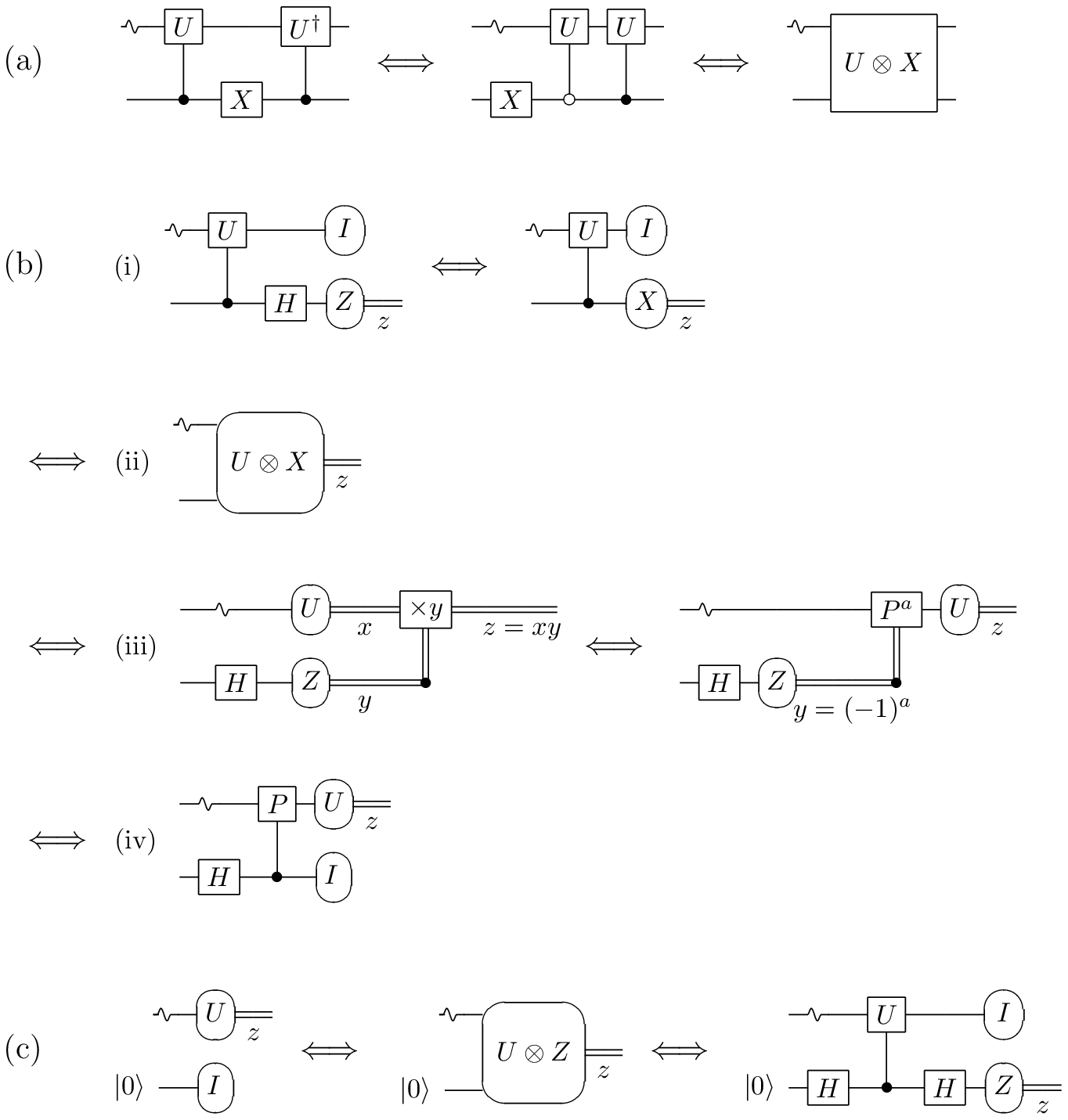}
\caption{The unitary operator $U$ is also Hermitian, i.e.,
$U=U^\dagger$, implying that all its eigenvalues are equal to $\pm1$.
(a)~Conjugation of $I\otimes X$ by a controlled-$U$ gives $U\otimes
X$.  In the middle circuit, $X$ has been pushed through the initial
controlled-$U$, turning the control on $\ket1$ to control on $\ket0$.
(b)~Circuit equivalences for disentangling circuits.  The first two
steps unitarily transform the measurement of $I\otimes Z$ to
$U\otimes X$. The next step reverts to a $Z$ measurement preceded by
a Hadamard and includes explicit classical control circuitry for
multiplying the outcomes of the $U$ and $Z$ measurements.  The fourth
step uses the bottom equivalence in Fig.~\protect\ref{fig5}(b) to
replace the post-measurement classical multiplication, which swaps
$\pm1$ outcomes, with a pre-measurement unitary $P$ that swaps the
$\pm1$ eigenspaces of $U$, i.e., $PUP^\dagger=-U$.  This step thus
requires that the $\pm1$ eigenspaces have the same dimensions. The
final step uses the principle of deferred measurement of
Fig.~\protect\ref{fig5}(c) to move the controlled operation through
the $Z$ measurement.  The labels (i)--(iv) are introduced for later
reference.  That (i)--(iv) all lead to the same probability for the
ultimate outcome~$z$ is worked out in detail in the text.
(c)~Measurement of $U$, achieved with the use of an ancillary qubit.
The first equivalence is true because the qubit's input state is
$\ket0$.  The second equivalence follows from using $Z=HXH$ twice,
surrounding an application of the equivalence in~(a). The result is a
standard method for measuring parity~$\Pi$.  Notice that this part
does not require that $U$ have $\pm1$ eigenspaces of the same
dimension. \label{fig6}}
\end{figure*}

We need several additional rules for handling measurements, the first
three of which, notation for a measurement whose outcome is
discarded, the rule for unitarily transforming measured observables,
and the principle of deferred
measurement~\cite{Griffiths1996b,Nielsen2000a}, are introduced and
described in Fig.~\ref{fig5}.  Figure~\ref{fig6} combines the
equivalences of Fig.~\ref{fig5} to derive a new set of equivalences,
spelled out in Fig.~\ref{fig6}(b), which lie at the heart of
interferometry.  These equivalences express a very important concept,
{\em disentanglement}, in several different ways.  If there is a most
important figure to master before considering Heisenberg-limited
interferometers, Fig.~\ref{fig6} is it.  In using the equivalences of
Fig.~\ref{fig6}(b), it is important to remember that they require
that the unitary operator $U$ be Hermitian, which means that it is an
observable all of whose eigenvalues are equal to $\pm1$, and that for
the last two circuits, the $\pm1$ eigenspaces of $U$ have the same
dimension.

The equivalence of Fig.~\ref{fig6}(a) is also used in
Fig.~\ref{fig6}(c) to derive a standard method for measuring a
Hermitian $U$, of which the parity of a field mode is an outstanding
example.  This measurement is founded on a technique called {\em
phase kickback\/}~\cite{Nielsen2000a,Cleve1999c}, which maps the
phase of $U$, $\pm1$ in this case, to a qubit.  Phase kickback is a
key element in many quantum algorithms.  This qubit-assisted method
for measuring the parity of a mode was proposed by Lutterbach and
Davidovich~\cite{Lutterbach1997a,Lutterbach1998a} and implemented in
Refs.~\cite{Nogues2000a,Bertet2002a}.

The equivalences in Fig.~\ref{fig6}(b), though easy to derive from
circuit equivalences, have wandered far enough down the road of
circuit diagrams that it is instructive to check that we have not
gone astray.  A good way to do this is to demonstrate the
equivalences in the conventional language of state vectors.  Suppose
that in Fig.~\ref{fig6}(b), the input state has the relative-state
decomposition
\begin{equation}
\ket\psi=\sqrt{q_0}\ket{\psi_0}\otimes\ket0+\sqrt{q_1}\ket{\psi_1}\otimes\ket1\;,
\label{eq:6in}
\end{equation}
where the states $\ket{\psi_0}$ and $\ket{\psi_1}$, though
normalized, are not necessarily orthogonal, and $q_0+q_1=1$.  We show
that circuits~(i)--(iv) of Fig.~\ref{fig6}(b) all lead to the same
probability for the outcome~$z$.

The controlled-$U$ in circuit~(i) transforms the input state to
\begin{equation}
\sqrt{q_0}\ket{\psi_0}\otimes\ket0+\sqrt{q_1}U\ket{\psi_1}\otimes\ket1\;,
\end{equation}
and the subsequent Hadamard gate, $H$, leaves the state
\begin{equation}
\frac{1}{\sqrt2}\Bigl(\sqrt{q_0}\ket{\psi_0}+\sqrt{q_1}U\ket{\psi_1}\Bigr)\otimes\ket0+
\frac{1}{\sqrt2}\Bigl(\sqrt{q_0}\ket{\psi_0}-\sqrt{q_1}U\ket{\psi_1}\Bigr)\otimes\ket1\;,
\end{equation}
from which it follows that the probability to obtain outcome $z$ in
the $Z$ measurement on the qubit~is
\begin{equation}
\label{eq:pzi}
p_z=\Bigl|\Bigl|\sqrt{q_0}\ket{\psi_0}+z\sqrt{q_1}U\ket{\psi_1}\Bigr|\Bigr|^2=
\frac{1}{2}\Bigl(1+2z\sqrt{q_0q_1}\,\mbox{Re}(\matrixel{\psi_0}{U}{\psi_1})\Bigr)\;.
\end{equation}

For circuit~(ii), the probability for outcome~$z$ follows from
the expectation value of $U\otimes X$ in the input state~(\ref{eq:6in}),
i.e.,
\begin{equation}
p_z=\frac{1}{2}
(1+z\matrixel{\psi}{U\otimes X}{\psi})
=\frac{1}{2}\Bigl(
1+z\sqrt{q_0q_1}\bigl(\matrixel{\psi_0}{U}{\psi_1}+\matrixel{\psi_1}{U}{\psi_0}\bigr)
\Bigr)\;,
\end{equation}
which is the same as Eq.~(\ref{eq:pzi}) because $U=U^\dagger$.

In circuit~(iii), the state after the initial Hadamard takes the form
\begin{equation}
\sqrt{r_{+1}}\frac{1}{\sqrt{2r_{+1}}}\Bigl(\sqrt{q_0}\ket{\psi_0}+\sqrt{q_1}\ket{\psi_1}\Bigr)\otimes\ket0+
\sqrt{r_{-1}}\frac{1}{\sqrt{2r_{-1}}}\Bigl(\sqrt{q_0}\ket{\psi_0}-\sqrt{q_1}\ket{\psi_1}\Bigr)\otimes\ket1\;,
\end{equation}
where
\begin{equation}
r_y=\Bigl|\Bigl|\sqrt{q_0}\ket{\psi_0}+y\sqrt{q_1}\ket{\psi_1}\Bigr|\Bigr|^2=
\frac{1}{2}\Bigl(1+2y\sqrt{q_0q_1}\,\mbox{Re}(\innerprod{\psi_0}{\psi_1})\Bigr)
\end{equation}
is the probability for the $Z$ measurement to yield outcome~$y$.  The
state of the mode after the $Z$ measurement yields outcome~$y$ is
\begin{equation}
\frac{1}{\sqrt{2r_y}}\Bigl(\sqrt{q_0}\ket{\psi_0}+
y\sqrt{q_1}\ket{\psi_1}\Bigr)\;.
\end{equation}
Hence, the conditional probability for the measurement of $U$ to have
outcome $x$, given outcome $y$ for the $Z$ measurement, is
\begin{align}
p_{x|y}&=\frac{1}{2}\Biggl(1+x
\biggl(\begin{matrix}\mbox{expectation value}\\\mbox{of $U$, given $y$}\end{matrix}
\biggr)\Biggr)\nonumber\\
&=\frac{1}{2}
\left(1+x\frac{1}{2r_y}
\Bigl(\sqrt{q_0}\bra{\psi_0}+y\sqrt{q_1}\bra{\psi_1}\Bigr)U
\Bigl(\sqrt{q_0}\ket{\psi_0}+y\sqrt{q_1}\ket{\psi_1}\Bigr)
\right)\nonumber\\
&=\frac{1}{2}\left(
1+\frac{x}{2r_y}
\Bigl(q_0\expect{\psi_0}{U}+q_1\expect{\psi_1}{U}\Bigr)
+\frac{xy\sqrt{q_0q_1}}{r_y}\mbox{Re}(\matrixel{\psi_0}{U}{\psi_1})
\right)\;.
\end{align}
The unconditioned probability for the outcome of the $U$ measurement is
\begin{equation}
p_x=\sum_y p_{x|y}r_y=
\frac{1}{2}\biggl(
1+x
\Bigl(q_0\expect{\psi_0}{U}+q_1\expect{\psi_1}{U}\Bigr)
\biggr)\;.
\end{equation}
In contrast, the conditional probability for the product of the two
outcomes, $z=xy$, is $p_{z|y}=p_{x=zy|y}$, which gives an
unconditioned probability
\begin{equation}
p_z=\sum_y p_{x=zy|y}r_y=
\frac{1}{2}\Bigl(
1+2z
\sqrt{q_0q_1}\,\mbox{Re}(\matrixel{\psi_0}{U}{\psi_1})
\Bigr)\;,
\end{equation}
the same as that in Eq.~(\ref{eq:pzi}).

In circuit~(iv), after the initial Hadamard and the controlled-$P$,
the state becomes
\begin{equation}
\frac{1}{\sqrt2}\Bigl(\sqrt{q_0}\ket{\psi_0}+\sqrt{q_1}\ket{\psi_1}\Bigr)\otimes\ket0+
\frac{1}{\sqrt2}P\Bigl(\sqrt{q_0}\ket{\psi_0}-\sqrt{q_1}\ket{\psi_1}\Bigr)\otimes\ket1\;.
\end{equation}
Hence, the probability of outcome $z$ in the measurement of $U$ is
\begin{align}
p_z
&=\frac{1}{2}
\biggl(1+z\frac{1}{2}
\Bigl(\sqrt{q_0}\bra{\psi_0}+\sqrt{q_1}\bra{\psi_1}\Bigr)U
\Bigl(\sqrt{q_0}\ket{\psi_0}+\sqrt{q_1}\ket{\psi_1}\Bigr)\nonumber\\
&\hphantom{\frac{1}{2}\Bigl(1+}
+z\frac{1}{2}
\Bigl(\sqrt{q_0}\bra{\psi_0}-\sqrt{q_1}\bra{\psi_1}\Bigr)PUP^\dagger
\Bigl(\sqrt{q_0}\ket{\psi_0}-\sqrt{q_1}\ket{\psi_1}\Bigr)
\biggr)\nonumber\\
&=\frac{1}{2}\Bigl(
1+2z
\sqrt{q_0q_1}\,\mbox{Re}(\matrixel{\psi_0}{U}{\psi_1})
\Bigr)\;,
\end{align}
where the final step uses that $P$ conjugates $U$ to its opposite,
i.e., $PUP^\dagger=-U$.  The final probability is again the same as
that in Eq.~(\ref{eq:pzi}).

\begin{figure*}
\center
\includegraphics{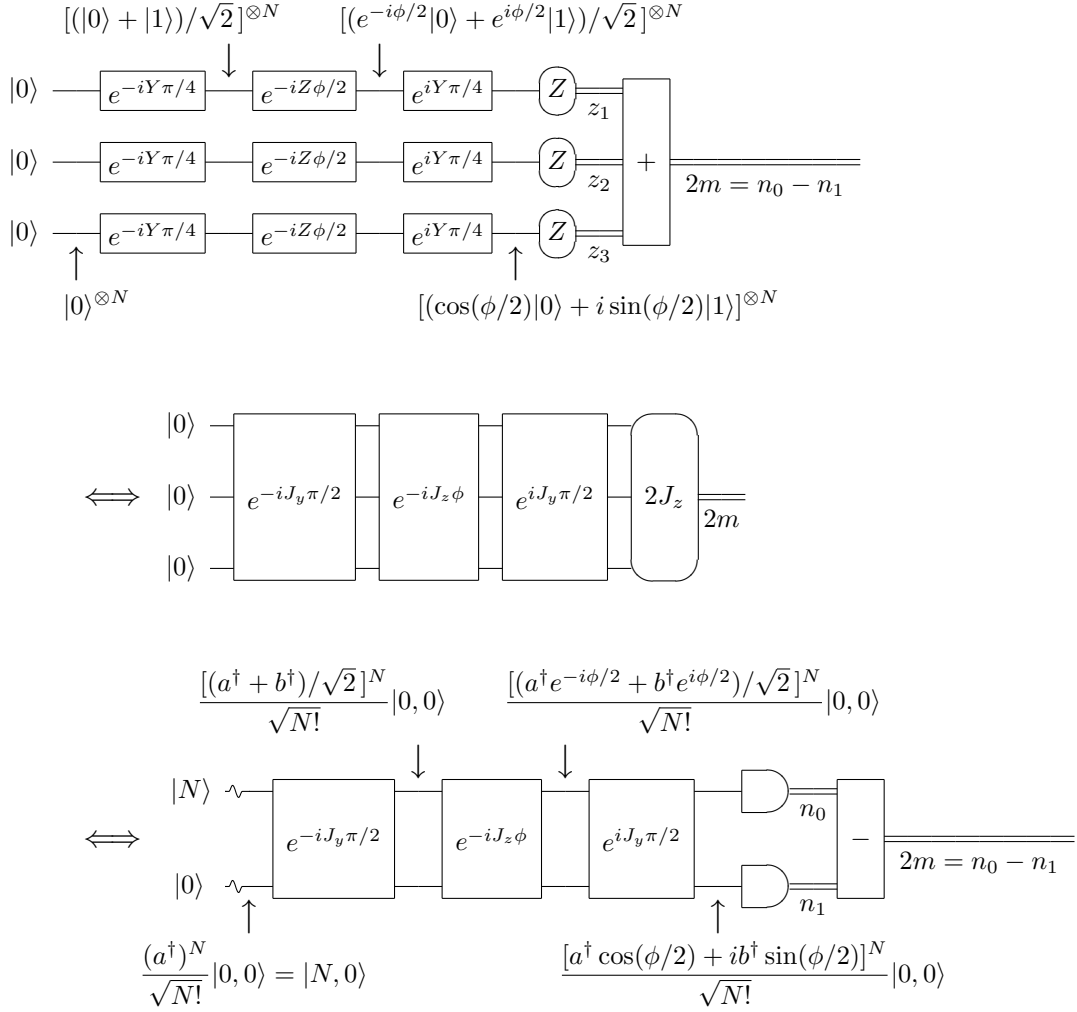}
\caption{Conventional qubit and equivalent modal interferometers,
which use independent particles.  The top circuit, a qubit
interferometer, clearly acts independently on the input qubits,
subjecting each to a~$\pi/2$ pulse, a rotation that imposes the phase
$\phi$, and a second~$\pi/2$ pulse.  The state of the $N$ qubits
after each gate is given.  Each qubit is measured in the standard
basis; the probabilities to be found in states $\ket0$ and $\ket1$
are $\cos^2(\phi/2)$ and $\sin^2(\phi/2)$. The $N$ outputs are
processed to extract the difference $2m=n_0-n_1$. The middle circuit
writes the same interferometer in terms of Schwinger operators, and
the bottom circuit completes the translation to a modal description.
The $N$ photons, initially all in mode~0, are subjected to a 50/50
beamsplitter, a phase shifter, and a second 50/50 beamsplitter, after
which the number of photons in each mode is counted, with the outputs
differenced to produce $2m=n_0-n_1$.  Quantum states are again
tracked after each gate.  The output $2m$ is binomially distributed;
the mean and variance of $m$ are $\langle
m\rangle=\frac{1}{2}N\cos\phi$ and $\Delta^2
m=\frac{1}{4}N\sin^2\!\phi$.  The resulting phase sensitivity,
$\delta\phi=\Delta m/|d\langle m\rangle/d\phi|=1/\sqrt N$, is the
quantum noise limit.  The overall interferometric transformation is a
rotation by $-\phi$ about the $x$ axis, i.e.,
$e^{iJ_y\pi/2}e^{-iJ_z\phi}e^{-iJ_y\pi/2}=e^{iJ_x\phi}$.
\label{fig7}}
\end{figure*}

There is an important special case of these four demonstrations,
typical of interferometric settings and instructive in figuring out
how these circuits work.  Suppose that all inputs to the circuit have
the form~(\ref{eq:6in}), with $\ket{\psi_0}$ and $\ket{\psi_1}$ being
orthogonal, but with a relative phase $\phi$ between $\ket{\psi_0}$
and $\ket{\psi_1}$.  Thus the inputs take the form
\begin{equation}
\sqrt{q_0}\ket{\psi_0}\otimes\ket0+\sqrt{q_1}e^{-i\phi}\ket{\psi_1}\otimes\ket1\;.
\label{eq:6in2}
\end{equation}
We can apply all our results for the three circuits simply by making
the replacement $\ket{\psi_1}\rightarrow e^{-i\phi}\ket{\psi_1}$. The
phase $\phi$, which can be thought of as the phase to be estimated in
an interferometer, is encoded in the entanglement of the upper wire
of the circuit with the qubit on the lower wire.  Furthermore,
suppose that $U$ swaps $\ket{\psi_0}$ and $\ket{\psi_1}$, i.e.,
$U\ket{\psi_0}=\ket{\psi_1}$ and $U\ket{\psi_1}=\ket{\psi_0}$.  If we
think of $\ket{\psi_0}$ and $\ket{\psi_1}$ as the standard states in
a two-dimensional Hilbert space, then $U$ is the Pauli $X$ operator
in this space, and $P$ can be chosen to be the Pauli $Z$ operator.

In this situation it is clear that circuit~(ii) employs a joint
measurement to read out the phase information directly.  In
circuit~(i), the role of the controlled-$U$ is to disentangle the
upper and lower wires, leaving the product state
$\ket{\psi_0}\otimes(\sqrt{q_0}\ket0+\sqrt{q_1}e^{-i\phi}\ket1)$, in
which the phase information is written on the qubit.  The subsequent
Hadamard transfers the phase information to amplitude information,
which determines the outcome probabilities for the measurement of
$Z$.  Likewise, in circuit~(iv), the state after the Hadamard and
controlled-$P$ is a product state,
$(\sqrt{q_0}\ket{\psi_0}+\sqrt{q_1}e^{-i\phi}\ket{\psi_1})\otimes(\ket0+\ket1)/\sqrt2$,
but in this case, the phase information is written onto the top line
of the circuit.  Circuits~(i) and~(iv) are thus both disentangling
circuits, which write the phase information onto one of the quantum
wires, where it can be read out by a measurement on that wire alone.
For both circuits, the probability of outcome $z$ is
$p_z=\frac{1}{2}(1+2z\sqrt{q_0q_1}\cos\phi)$.

\begin{figure*}
\center
\includegraphics{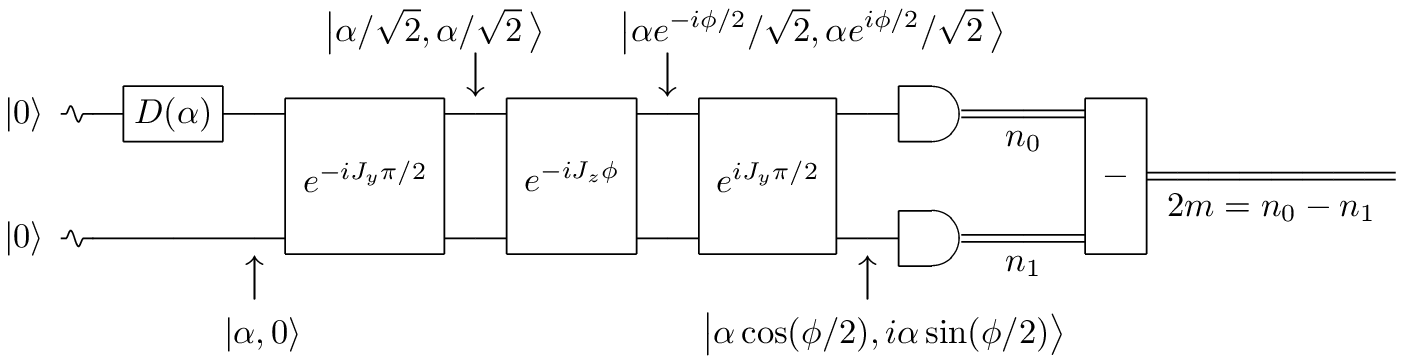}
\caption{Coherent-state version of a conventional modal
interferometer.  The input coherent state
$\ket{\alpha,0}=D(\alpha)\otimes
I\ket{0,0}=e^{-|\alpha|^2/2}\sum_{N=0}^\infty(\alpha^N/\sqrt{N!}\,)\ket{N,0}$
is the result of applying the displacement operator $D(\alpha)$ to
the vacuum state of mode~0.  The displacement operator is the
mathematical description of the action of a classical driving field
on the mode; the action of an ideal laser operating far above
threshhold is well approximated by a displacement operator, apart
from a slow drift in phase.  Since the interferometer itself is
number-preserving, the different $N$-photon sectors of the coherent
state are processed independently, implying that the output
probabilities are obtained by averaging the binomial distribution of
Fig.~\ref{fig7} over the coherent state's Poisson distribution for
total photon number.  The resulting mean and variance of $m$ are
$\langle m\rangle=\frac{1}{2}|\alpha|^2\cos\phi$ and $\Delta^2
m=\frac{1}{4}|\alpha|^2\sin^2\!\phi$, which gives phase sensitivity
$\delta\phi=1/|\alpha|$.  We can think of the coherent-state
interferometer as a classical interferometer contaminated by the shot
noise of the coherent state's Poisson distribution.  The classicality
of the coherent states appears in the state description, tracked
through the gates, as the fact that the phase $\phi$ appears not as a
relative phase between states in superposition, but as a phase change
of coherent-state (classical) complex amplitudes. \label{fig8}}
\end{figure*}

Circuit~(iii) is also disentangling, but works in a different way.
With the current assumptions, the two outcomes of the measurement of
$Z$ are equally likely, i.e., $r_y=1/2$, and the state of the upper
wire after the measurement of $Z$ yields outcome~$y$ is
$\sqrt{q_0}\ket{\psi_0}+y\sqrt{q_1}e^{-i\phi}\ket{\psi_1}$.  The
upper wire now carries the phase information, but one needs the
outcome~$y$ to know how to extract phase information from the
subsequent measurement of $U$.  The conditional probability for the
outcome~$x$ of the $U$ measurement, given $y$, is
$p_{x|y}=\frac{1}{2}(1+2xy\sqrt{q_0q_1}\cos\phi)$.  If one does not
use the outcome $y$, then the phase produces no interference, the
unconditioned probability for $x$ being $p_x=1/2$, but if one does
use the phase information, as in the classical control of
circuit~(iii), the phase does produce interference, the unconditioned
probability for the product, $z=xy$, being the same as for
circuits~(i), (ii), and~(iv).

In circuits~(i) and~(iv), disentanglement occurs as a consequence of
coherent operations, but in circuit~(iii), disentanglement is a
consequence of the measurement of $Z$.  This measurement-induced (or
post-selected) disentanglement was originally called a {\em quantum
eraser\/}~\cite{Scully1981a,Scully1991a,Walborn2002a}; the modern
understanding is that what is ``erased'' is the entanglement that
destroys interference when one measures only the top wire or the
bottom wire.  A nice feature of the circuit~(iii) is that one can
regard the Hadamard and measurement of $Z$ as occurring either before
or after the measurement of $U$.  In the former case, they seem to
prepare a state on the top wire that carries the phase information
and in the latter case, they seem to ``erase'' the entanglement that
destroyed the interference on the top wire, but there is no
difference between these two cases, only a difference in perspective.

\section{Quantum-noise-limited interferometers}
\label{sec:qnlint}

The chief purpose of this paper is to examine and relate different
methods for achieving Heisenberg-limited sensitivity, i.e., phase
sensitivity that scales as $1/N$ with the number $N$ of atoms or
photons.  Before taking up that task in Sec.~\ref{sec:Fockstateint},
however, we first look, in Fig.~\ref{fig7}, at conventional qubit and
modal interferometers, which use independent particles.  Qubit
interferometers are typically atomic (Ramsey) interferometers, and
modal interferometers employ two modes of the electromagnetic field.
The sensitivity of a conventional interferometer scales as $1/\sqrt
N$, the so-called {\em quantum noise limit\/} (or shot-noise
limit)~\cite{Bachor2004a}.

\begin{figure*}
\center
\includegraphics[width=6.4in]{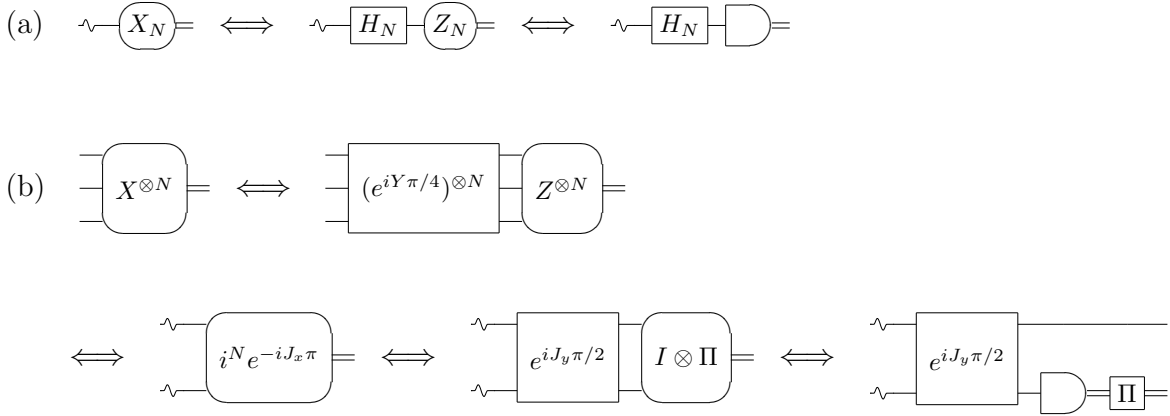}
\caption{(a)~A measurement of $X_N=\outerprod{0}{N}+\outerprod{N}{0}$
is equivalent to an effective Hadamard gate $H_N=(X_N+Z_N)/\sqrt2$
followed by a measurement of $Z_N=\proj0-\proj N$.  A subscript~$N$
on a two-dimensional unitary operator signifies that operator acting
in the ``qubit'' modal subspace spanned by standard states $\ket0$ and
$\ket N$.  The final form, which replaces the measurement of $Z_N$ by
photocounting, is valid if the input state lies in the ``qubit''
subspace spanned by $\ket0$ and $\ket N$; it is an example of
extending a measurement outside a subspace of interest.  (b)~A
measurement of $X^{\otimes N}$ on $N$ qubits is equivalent to
measuring $i^Ne^{-iJ_x\pi}$ on modes, and this in turn is equivalent
to a 50/50 beamsplitter followed by a measurement of the parity of
mode~1. Notice that $X^{\otimes N}=i^Ne^{-iJ_x\pi}$ is the qubit FLIP
and modal SWAP operator:
$i^Ne^{-iJ_x\pi}\ket{n_0,n_1}=\ket{n_1,n_0}$.  We use this
measurement in our discussion of N00N-state interferometers, where
all that is really needed is a measurement of the observable
$\outerprod{N,0}{0,N}+\outerprod{0,N}{N,0}$ in the two-dimensional
subspace spanned by $\ket{N,0}$ and $\ket{0,N}$; $i^Ne^{-iJ_x\pi}$ is
a natural extension of this observable into the entire Hilbert space.
Throughout this paper, we symbolize a parity measurement as in the
last circuit---photocounting followed by classical extraction of
parity---but the reader should remember that
Fig.~\protect\ref{fig6}(c), with $U=\Pi$, provides another, standard
circuit for measuring parity, by mapping the parity to an ancillary
qubit.\label{fig9}}
\end{figure*}

\begin{figure*}
\center
\includegraphics[width=6.5in]{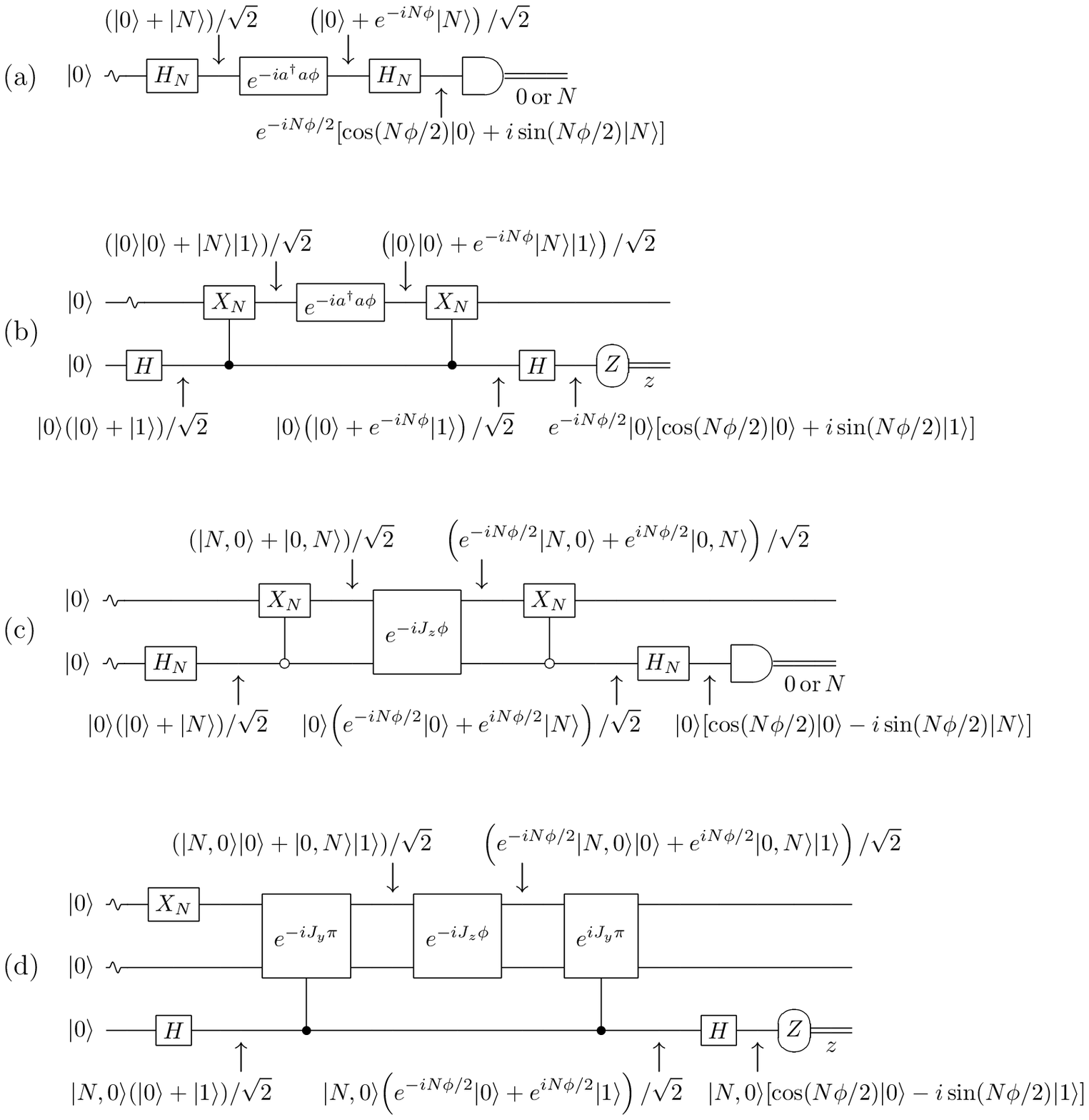}
\caption{Four Heisenberg-limited phase-estimation protocols that use
Fock-state superpositions: (a)~single-mode phase estimation;
(b)~single-mode, qubit-assisted phase estimation, (c)~two-mode
interferometer, called a N00N interferometer after the state before
the phase shifter; (d)~two-mode, qubit-assisted interferometer. In
the qubit-assisted protocols, the qubit can be thought of as a second
or third arm of an interferometer.  Quantum states are tracked
through each protocol.  In protocols~(b) and~(d), inserting a gate
$e^{i\theta/2}e^{-iZ\theta/2}$ between the two controls on the qubit
wire shifts the fringe pattern by angle $\theta$; in particular, an
$S$ gate shifts the fringes by $\pi/2$.  In protocols~(a) and~(c),
the output is either 0~photons or $N$~photons; mapping 0 to $+1$ and
$N$ to $-1$, the outcome becomes $z=\pm1$, as in protocols~(b)
and~(d).  In all cases, the probabilities of the two outcomes are
$p_z=\frac{1}{2}(1+z\cos N\phi)$.  The mean and variance of $z$,
$\langle z\rangle=\cos N\phi$ and $\Delta^2 z=\sin^2\!N\phi$, give
Heisenberg-limited phase sensitivity $\delta\phi=\Delta z/|d\langle
z\rangle/d\phi|=1/N$.  The photocounting in the N00N
interferometer~(c) can be replaced by a measurement of $Z_N$. Then
the operations after the phase shifter are equivalent to a
measurement of $X_N\otimes X_N$, which for the N00N-state input, is
the same as a measurement of
$\outerprod{N,0}{0,N}+\outerprod{0,N}{N,0}$; this, in turn, as
pointed out in Fig.~\protect\ref{fig9}(b), is equivalent on
N00N-state inputs to a measurement of the modal SWAP operator
$i^Ne^{-iJ_x\pi}$. \label{fig10}}
\end{figure*}

A couple of aspects of the interferometers in Fig.~\ref{fig7} deserve
attention.  In the modal version, the input Fock state $\ket{N}$ for
mode~0 can be replaced by a coherent state
$\ket{\alpha}=D(\alpha)\ket0$, where $D(\alpha)=e^{\alpha
a^\dagger-\alpha^* a}$ is the modal displacement operator.  It is
worth recording that $e^{-i\bm{J\cdot\hat n}\theta}D(\alpha)\otimes
D(\beta)e^{i\bm{J\cdot\hat n}\theta}=D(\alpha')\otimes D(\beta')$,
where
\begin{equation}
\left(\,\begin{matrix}\alpha'\\\beta'\end{matrix}\,\right)\!
=M\!\left(\,\begin{matrix}\alpha\\\beta\end{matrix}\,\right)\!\;,
\end{equation}
with $M$ given by Eq.~(\ref{eq:M}).  This implies that
\begin{equation}
e^{-i\bm{J\cdot\hat n}\theta}\ket{\alpha,\beta}=\ket{\alpha',\beta'}\;.
\end{equation}
A coherent-state interferometer is depicted and described in
Fig.~\ref{fig8}.  It is a semi-realistic rendition of what is done in
optical interferometry, since an ideal laser produces a very good
approximation to a coherent state.

It is clear from both the qubit and modal versions of a conventional
interferometer that the output is insensitive to common-mode phase
shifts.  This is the primary advantage of measuring phase shifts in
an interferometric setting.  It is also clear, however, that the
equatorial axes of the two 50/50 beamsplitters (or $\pi/2$
transitions) must be well defined relative to one another.  A change
in the equatorial axis of the second beamsplitter introduces a
relative phase shift across the beamsplitter that is
indistinguishable from the phase $\phi$.  In most applications, where
one is only interested in changes in $\phi$, the actual requirement
is that the relative phase shifts across the two beamsplitters be
stable relative to one another over times somewhat longer than the
time scale of changes in $\phi$.

\begin{figure*}
\center
\includegraphics[width=6.4in]{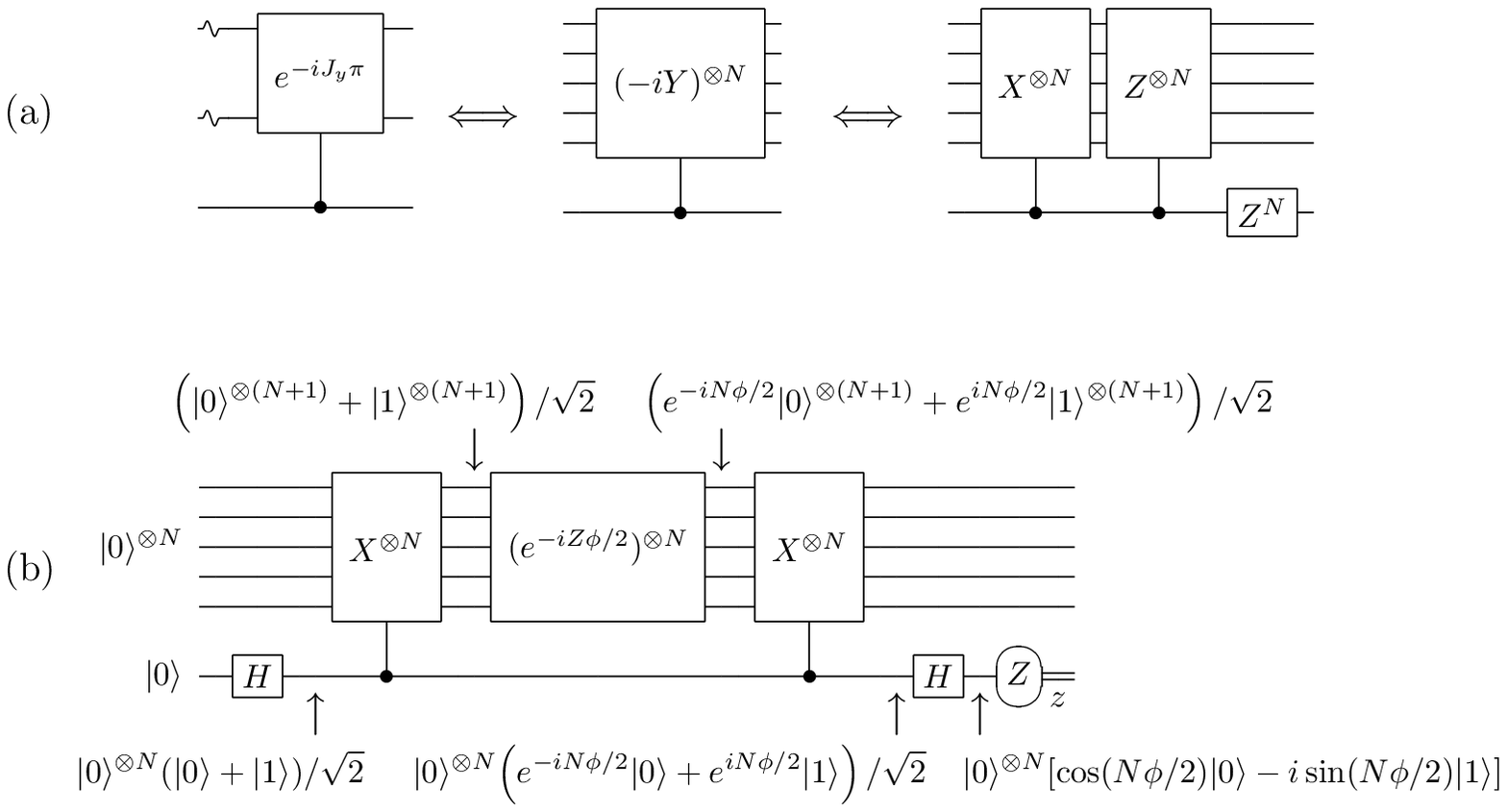}
\caption{(a)~Circuit equivalences that equate a controlled-SWAP
between two modes to controlled-NOTs and controlled-SIGNs on $N=5$
qubits, plus a $Z^N$ on the control qubit.  (b)~$N$-qubit cat-state
interferometer, obtained by applying~(a) to the two-mode,
qubit-assisted interferometer in Fig.~\protect\ref{fig10}(d).  The
interferometer is named after the cat
state~$(\ket0^{\otimes(N+1)}+\ket1^{\otimes(N+1)})/\sqrt2$ at the
input to the phase shifters; it has the same outcome probabilities
and phase sensitivity as Fig.~\protect\ref{fig10}(d).  Proposals for
cat-state interferometers often also apply the phase shift to the
control qubit; that does not happen here because of the special role
of the control qubit, which is inherited from the two-mode,
qubit-assisted circuit in Fig.~\protect\ref{fig10}(d). \label{fig11}}
\end{figure*}

In different implementations, the stability requirements between the
two beamsplitters assume different forms.  In optical-frequency
interferometers, where the modes typically correspond to different
spatial paths or different polarizations, the phase shifts across the
beamsplitters are determined by coatings on the beamsplitters and by
placement or orientation of the beamsplitters; it is relatively easy
to keep these stable.  In atomic interferometers, the relative phase
shifts across the $\pi/2$ pulses are chiefly determined by the times
at which the pulses are applied; it is not difficult to keep the time
separating the two pulses stable at microwave frequencies.

\section{Fock-state-superposition interferometers}
\label{sec:Fockstateint}

Entanglement among particles allows one to improve the sensitivity
scaling for estimating a phase beyond that offered by the quantum
noise limit.  The optimal sensitivity scaling, known as the {\em
Heisenberg limit}, is given by
$1/N$~\cite{Giovannetti2006a,Boixo2007a,Boixo2009au}.  There are many
different methods for achieving this limit, most of which involve
using superpositions of two Fock states of one or two modes.  In this
section we consider four such methods and use circuit equivalences to
show they are equivalent.  In Sec.~\ref{sec:coherentstateint}, we
turn to corresponding methods that use superpositions of coherent
states and find that they, too, saturate the Heisenberg limit, but
with considerably better prospects for realization in the laboratory.

Before getting started on Fock-state-superposition interferometers,
we introduce two measurement rules in Fig.~\ref{fig9}.  The first, in
Fig.~\ref{fig9}(a), is useful in our discussion of phase estimation
using a single mode, and the second, in Fig.~\ref{fig9}(b), comes
into play when we discuss genuine interferometers, which use two
modes, such as N00N-state interferometers.  In the first rule, we
introduce the notation that a subscript~$N$ on a two-dimensional
unitary operator denotes that operator acting in the ``qubit'' modal
subspace spanned by standard states $\ket0$ and $\ket N$.  The second
rule is important for interferometers because it converts a
measurement of the modal SWAP operator into a measurement that might
be doable, i.e, a measurement of the parity of one mode behind a
50/50 beamsplitter.

The ability to convert to a parity measurement---and its
importance---has been appreciated in many
papers~\cite{Gerry2000a,Gerry2001a,Gerry2002a,Campos2003a,Campos2005a,Gerry2007a,Chiruvelli2009au,YGao2009au}.
Parity can be measured by photocounting followed by classical
extraction of the parity---this is how we represent a parity
measurement throughout this paper---but it can also be measured using
the protocol of Fig.~\ref{fig6}(c) with $U=\Pi$, which maps the
parity to an ancillary qubit, and by other methods that rely on the
interference of two
modes~\cite{Wallentowitz1996a,Banaszek1996a,Banaszek1999b}.  Indeed,
these methods can be applied more generally to measuring the
displaced parity operator, a measurement that comes up naturally in
our discussion in Sec.~\ref{sec:coherentstateint}, and that is the
basis for direct measurements of the Wigner function of a
mode~\cite{Lutterbach1997a,Lutterbach1998a,Nogues2000a,Bertet2002a,%
Hofheinz2009a,Wallentowitz1996a,Banaszek1996a,Banaszek1999b}.

\begin{figure*}
\center
\includegraphics{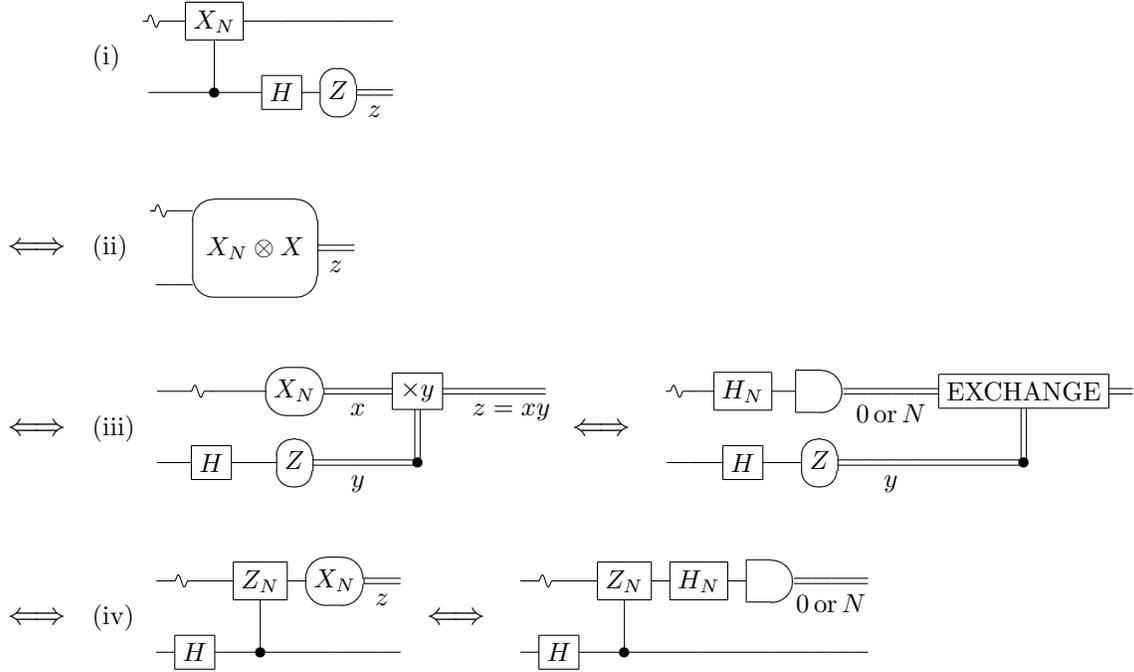}
\caption{Circuit equivalences for disentangling circuits when the
unitary $U$ of Fig.~\protect\ref{fig6}(b) is $X_N$.  Circuits are
numbered as in Fig.~\protect\ref{fig6}(b). Circuits~(iii) and~(iv)
are converted to more useful forms by replacing the measurement of
$X_N$ by an effective Hadamard followed by photocounting, as in
Fig.~\protect\ref{fig9}(a).  In~(iii), the classical
controlled-EXCHANGE swaps the outcomes 0 and $N$ when the outcome of
the $Z$ measurement is $y=+1$.  In~(iv), $Z_N$ is the operator that
conjugates $X_N$ to its opposite.  \label{fig12}}
\end{figure*}

\begin{figure*}
\center
\includegraphics{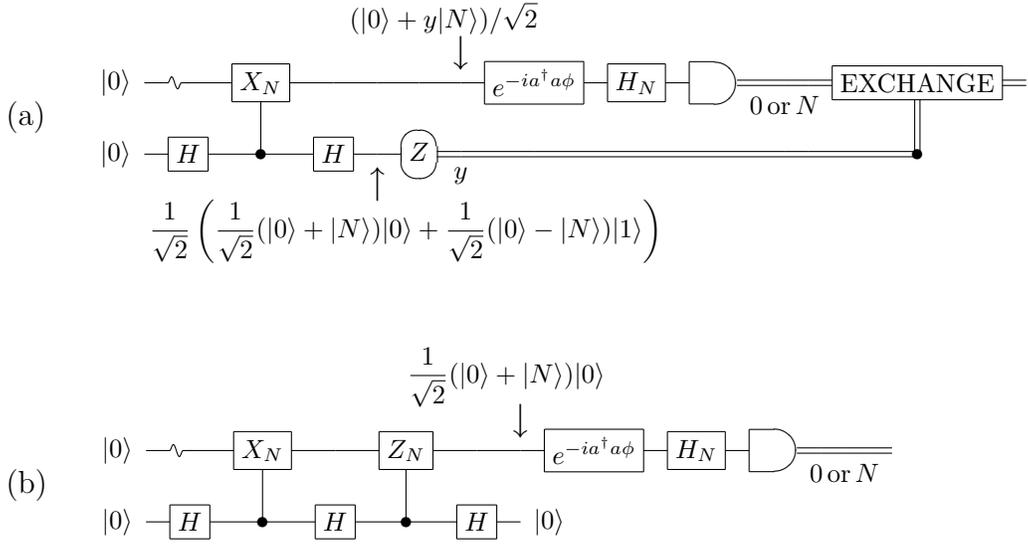}
\caption{(a)~Version of the circuit of Fig.~\protect\ref{fig10}(a)
obtained by applying the second form of circuit~(iii) in
Fig.~\protect\ref{fig12} to the circuit of
Fig.~\protect\ref{fig10}(b).  In this version, the superposition
state input to the phase shifter is prepared by post-selection based
on the outcome of the $Z$ measurement on the qubit.  The two possible
outcomes occur with equal probability; the two input states,
$(\ket0\pm\ket N)/\sqrt2$, produce fringe patterns that are $\pi$ out
of phase.  Averaging over outcomes destroys the fringe pattern, but
the classical control circuitry restores a single fringe pattern. The
initial circuitry and qubit measurement entangle and then disentangle
the mode and the qubit; if the qubit measurement is regarded as
occurring after the photocounting, this is the action of a quantum
eraser.  (b)~Version of the circuit of Fig.~\protect\ref{fig10}(a)
obtained by applying the second form of circuit~(iv) in
Fig.~\protect\ref{fig12} to the circuit of
Fig.~\protect\ref{fig10}(b).  Since $Z_N$ commutes with $a^\dagger
a$, the controlled-$Z_N$ passes unscathed through the phase shifter,
becoming part of a coherent procedure for the superposition state
input to the phase shifter.  A final Hadamard gate is added to the
qubit wire to return the qubit to its initial state $\ket0$.
\label{fig13}}
\end{figure*}

Figure~\ref{fig10} depicts circuit diagrams, labeled (a)--(d), for
four phase-estimation protocols that employ superpositions of two
Fock states. It is clear from the tracking of quantum states through
these protocols that they are closely related and that all achieve
Heisenberg-limited sensitivity by having a fringe pattern that
oscillates $N$ times as fast as the independent-particle fringe
pattern of a conventional interferometer. We emphasize that these
methods differ from conventional interferometry in that they are not
committed to placing beamsplitters immediately before and after the
phase shifter, as in a conventional interferometer, but rather allow
any procedure for preparing the input state to the phase shifter and
any measurement procedure after the \hbox{beamsplitter}.

Of the four protocols in Fig.~\ref{fig10}, the best known is
interferometer~(c), which inputs N00N states,
$(\ket{N,0}+\ket{0,N})/\sqrt2$, to the differential phase shifter
$e^{-iJ_z\phi}$ and thus can be called a N00N
interferometer~\cite{Gerry2000a,Boto2000a,Gerry2001a,HLee2002a,Gerry2002a,Campos2003a,Gerry2007a,Dowling2008a}.
We remind the reader that the two-mode, qubit-assisted
interferometer~(d) is equivalent to an $N$-qubit cat-state
interferometer~\cite{Bollinger1996a,Huelga1997a,Shaji2007a}, as is
shown in Fig.~\ref{fig11}.

\begin{figure*}
\center
\includegraphics[width=6.4in]{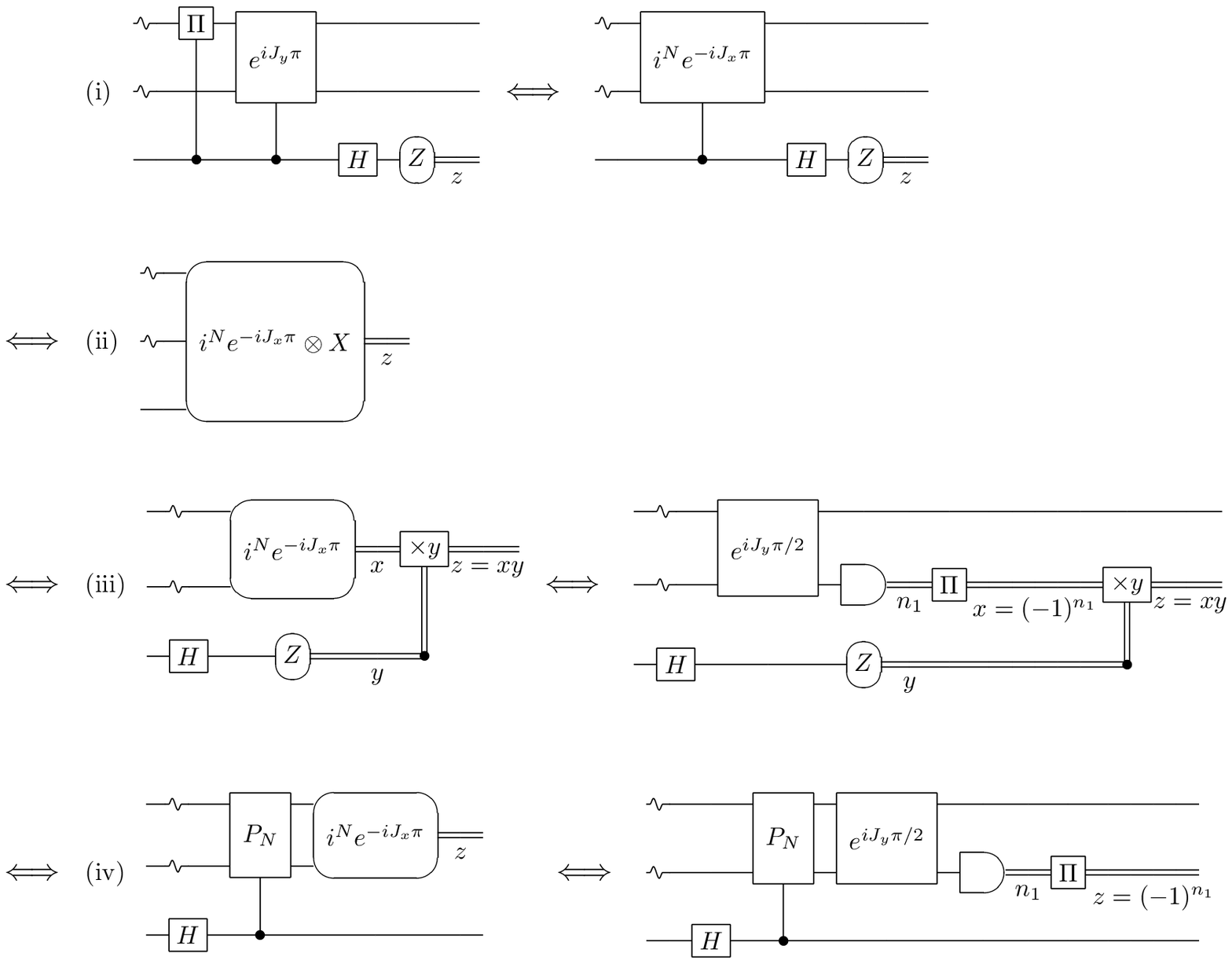}
\caption{Circuit equivalences for disentangling circuits when the
unitary $U$ of Fig.~\protect\ref{fig6}(b) is $e^{iJ_y\pi}\Pi\otimes
I=i^NY^{\otimes N}(-Z)^{\otimes N}=X^{\otimes N}=i^Ne^{-iJ_x\pi}=
e^{-iJ_y\pi/2}I\otimes\Pi e^{iJ_y\pi/2}$.  The
controlled-($\Pi\otimes I$) is included to make $U$ Hermitian as well
as unitary.  Circuits are numbered as in Fig.~\protect\ref{fig6}(b).
Circuits~(iii) and~(iv) are converted to more useful forms by
replacing the measurement of $i^Ne^{-iJ_x\pi}$ by a 50/50
beamsplitter followed by photocounting on mode~1 and classical
extraction of parity, as in Fig.~\protect\ref{fig9}(b).  The step to
circuit~(iv) is not generally valid, since it requires that the
$\pm1$ eigenspaces of $i^Ne^{-iJ_x\pi}$ have the same dimension,
which is only true when $N$ is odd.  We use circuit~(iv), however,
only when the modal inputs are confined to the subspace spanned by
$\ket{N,0}$ and $\ket{0,N}$; in this case we can define $P_N$ by
$P_N\ket{N,0}=\ket{N,0}$ and $P_N\ket{0,N}=-\ket{0,N}$.  When $N$ is
odd, we can choose $P_N=I\otimes\Pi$. \label{fig14}}
\end{figure*}

The easiest equivalence to show is that of circuits~(b) and~(c).
Insert a $Z$ at the beginning of the qubit wire in~(b).  Push it to
the end of the circuit through the measurement, thereby converting
the controlled operations to control on $\ket0$.  Replace the qubit
with the two-dimensional modal subspace spanned by standard states
$\ket0$ and $\ket N$.  Insert an irrelevant global phase
$e^{iN\phi/2}$; this changes the phase shift $e^{-ia^\dagger a\phi}$
on mode~0 to a differential phase shift $e^{-iJ_z\phi}$ and completes
the conversion of~(b) to~(c).

The other equivalences, (b) to~(a) and~(d) to ~(c), are more
interesting.  It is easy to see that circuit~(b) is a version of~(a)
with the phase information carried by an entangled state of the mode
and the qubit; likewise, circuit~(d) is a version of~(c) with the
phase information carried by entanglement of the two modes with the
qubit.  In the qubit-assisted protocols, (b) and~(d), the
post-phase-shift operations write all the phase information onto the
qubit by disentangling it from the mode(s), whereas in~(a) and~(c),
the phase information is on the mode(s).  The key to showing the
equivalences is thus to use the disentangling equivalences of
Fig.~\ref{fig6}(b) to move the phase information in~(b) and ~(d) onto
the mode(s) instead of onto the qubit.

\begin{figure*}
\center
\includegraphics[width=6.4in]{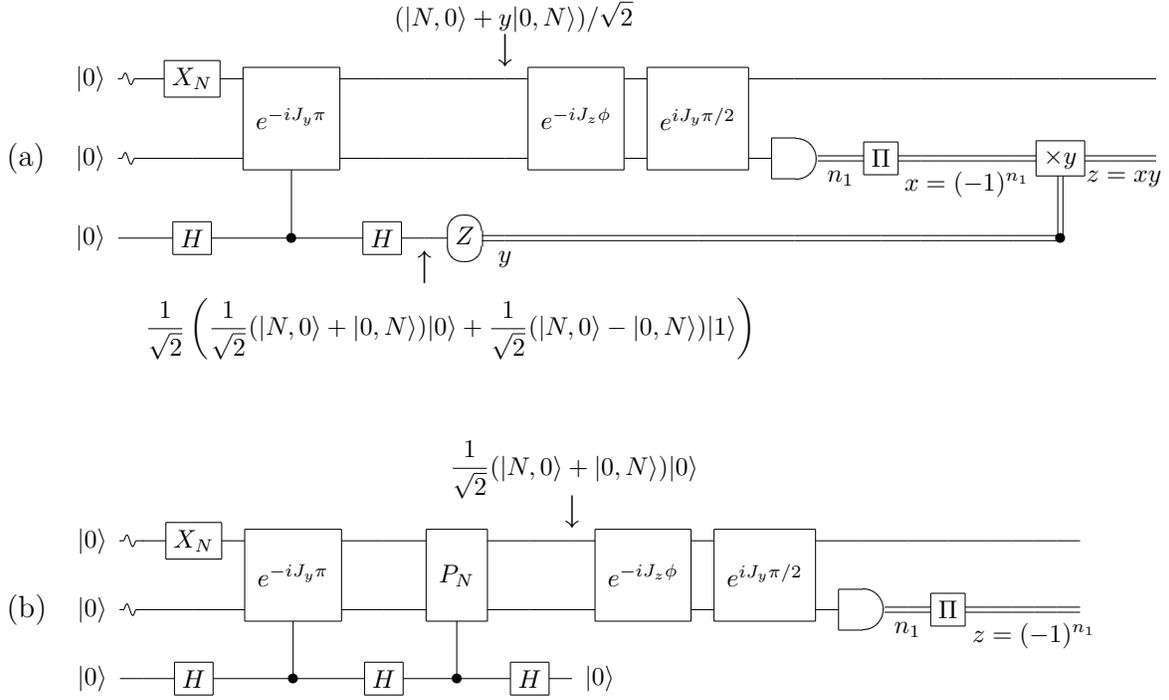}
\caption{(a)~Version of the N00N circuit of
Fig.~\protect\ref{fig10}(c) obtained by applying the second form of
circuit~(iii) in Fig.~\protect\ref{fig14} to the circuit of
Fig.~\protect\ref{fig10}(d).  The controlled-($I\otimes\Pi$) at the
beginning of Fig.~\protect\ref{fig14}(i) is moved to the beginning of
Fig.~\protect\ref{fig14}(iii); it is pushed through the phase shifter
and then through the controlled-$e^{-iJ_y\pi}$, becoming a
controlled-($I\otimes\Pi$) [because $\Pi\otimes I
e^{-iJ_y\pi}=(-Z)^{\otimes N}(-iY)^{\otimes N}=(-iY)^{\otimes
N}Z^{\otimes N}=e^{-iJ_y\pi}I\otimes\Pi$], which can be omitted on
mode~1's vacuum input.  In this version, the superposition state
input to the phase shifter is prepared by post-selection based on the
outcome of the $Z$ measurement on the qubit.  The two possible
outcomes occur with equal probability; the two N00N inputs,
$(\ket{N,0}\pm\ket{0,N})/\sqrt2$, produce fringe patterns that are
$\pi$ out of phase.  Averaging over outcomes destroys the fringe
pattern, but the classical control circuitry restores a single fringe
pattern.  The initial circuitry and qubit measurement entangle and
then disentangle the modes and the qubit; if the qubit measurement is
regarded as occurring after the photocounting, this is the action of
a quantum eraser.  (b)~Version of the N00N circuit of
Fig.~\protect\ref{fig10}(c) obtained by applying the second form of
circuit~(iv) in Fig.~\protect\ref{fig14} to the circuit of
Fig.~\protect\ref{fig10}(d).  The controlled-($\Pi\otimes I$) is
handled as in~(a).  Since $P_N$ commutes with $J_z$, the
controlled-$P_N$ passes through the phase shifter, becoming part of a
coherent procedure for the superposition state input to the phase
shifter.  A final Hadamard gate is added to the qubit wire to return
the qubit to its initial state $\ket0$.  The measurement procedure
in~(a) and~(b)---50/50 beamsplitter followed by photocounting and
classical extraction of parity---doesn't look like the measurement
procedure in Fig.~\protect\ref{fig10}(c), but both procedures are
equivalent to measuring the modal SWAP operator, $i^Ne^{-iJ_x\pi}$,
as noted in Figs.~\protect\ref{fig9} and~\protect\ref{fig10}.
\label{fig15}}
\end{figure*}

Figures~\ref{fig12}--\ref{fig15} demonstrate the two equivalences;
the demonstration that~(d) is equivalent to~(c) is very closely
analogous to the demonstration that~(b) is equivalent to~(a).
Figures~\ref{fig12} and \ref{fig14} specialize the disentangling
circuits of Fig.~\ref{fig6}(b) to the situations in circuits~(b)
and~(d), and Figs.~\ref{fig13} and~\ref{fig15} apply these to
circuits~(b) and~(d) to convert them to versions of~(a) and~(c).  The
converted circuits have particular procedures, assisted by the qubit,
for preparing an appropriate input state to the phase shifter.  These
preparation procedures are either post-selected, i.e., based on the
outcome of a measurement on the qubit, or completely coherent.  They
require the ability to make $N$-photon states and to control
two-dimensional Fock-state subspaces, both of which are very hard to
do experimentally.  Attempts to make N00N states beyond $N=2$ up till
now have had to be nondeterministic~\cite{Dowling2008a}.

On the other hand, the superposition state that powers~(a),
$(\ket0+\ket N)/\sqrt2$, has been made deterministically for $N=1$
to~5 in a qubit/cavity setting, where the qubit is a superconducting
circuit and the mode belongs to a microwave-frequency
resonator~\cite{Hofheinz2009a}.  The state-preparation protocol used
in that experiment, due to Law and Eberly~\cite{Law1996a}, can be
used in principle to create arbitrary superpositions of resonator
Fock states.  It alternates rotations of the qubit with carefully
timed Rabi couplings of the qubit to the resonator to deposit
successive photons in the resonator mode.  This protocol does not
lend itself readily to the circuit notation of this paper, nor in its
reliance on a ladder of two-level transitions, is it an efficient
method for making large-$N$ superpositions $(\ket0+\ket N)/\sqrt2$.

Gerry and Campos~\cite{Gerry2001a} have proposed a realization of the
post-selected preparation procedure of Fig.~\ref{fig15}(a). The role
of the qubit is taken over by two field modes that carry a single
photon, and the controlled-$e^{-iJ_y\pi}$ is achieved by a cross-Kerr
coupling, which does a controlled swap of the coupled modes.

\section{Coherent-state-superposition interferometers}
\label{sec:coherentstateint}

Just as for conventional interferometers, we can turn a
Heisenberg-limited interferometer from one that uses superpositions
of Fock states to a corresponding one that uses superpositions of
coherent states.  In this section, we take the single-mode and
two-mode, qubit-assisted protocols of Fig.~\ref{fig10}(b) and~(d) and
convert them to run on coherent states.  We then apply disentangling
measurement equivalences to convert to protocols that prepare input
states to the phase shifter that are superpositions of two coherent
states.  For a single mode, a superposition of two coherent states is
often called a cat state, but we refer to it as a {\em modal cat
state\/} to distinguish it from the cat state of $N$ qubits.

Versions of our preparation procedures for superpositions of coherent
states turn out to be practical, at least at microwave frequencies.
Ultimately, there are two features that make coherent superpositions
more practical than Fock-state superpositions.  First, coherent
states have their own phase, which we can play with in addition to
the relative phase in superpositions; second, coherent states can be
created and manipulated by coherent driving fields, represented by
the displacement operator, thus providing an additional, crucial
transformation that is not useful when dealing with Fock states.

\subsection{Single-mode coherent-state-superposition phase estimation}
\label{subsec:onecss}

\begin{figure*}
\center
\includegraphics[width=6.4in]{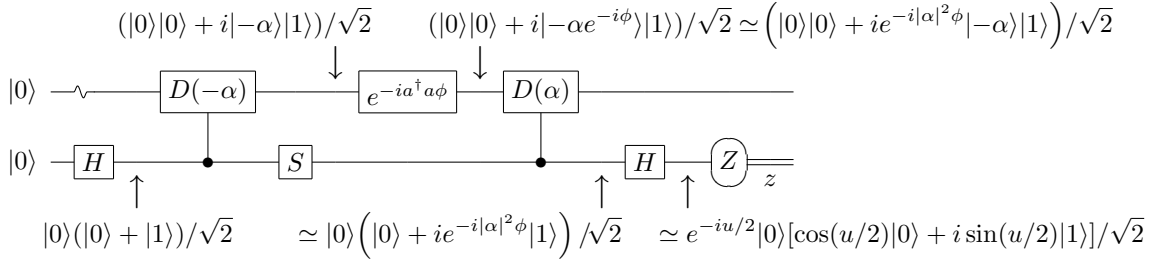}
\caption{Coherent-state analogue of the single-mode, qubit-assisted
phase-estimation protocol in Fig.~\protect\ref{fig10}(b).  In
constructing the analogue, one replaces the two controlled-$X_N$s
with controlled displacements that are inverses of one another.  Here
we make the first a controlled-$D(-\alpha)$ and the second a
controlled-$D(\alpha)$.  We also introduce an $S$ gate on the qubit
to shift the fringe pattern by $\pi/2$ so that optimal sensitivity
occurs at $\phi=0$; this fringe shift turns out to fit naturally into
further manipulations of the circuit.  Quantum states are tracked
through the circuit.  An approximately equal sign ($\simeq$), here
and henceforth, indicates an expression that assumes small phase
shifts, $|\alpha|^2\phi^2\ll|\alpha|^2\phi\ll1$.  The approximate
expressions emphasize the similarity to Fig.~\protect\ref{fig10}(b);
exact expressions are given in the text.  In the final state, we use
the variable $u=|\alpha|^2\phi-\pi/2$ for brevity; it expresses the
$\pi/2$ phase shift produced by the $S$ gate.  For small phase
shifts, the probability of outcome~$z$ is
$p_z\simeq\frac{1}{2}[1+z\sin(|\alpha|^2\phi)]$, which gives
Heisenberg-limited sensitivity $\delta\phi=1/|\alpha|$.
\label{fig16}}
\end{figure*}

Figure~\ref{fig16} gives the coherent-state analogue of the
single-mode, qubit-assisted protocol of Fig.~\ref{fig10}(b).  The
state after the controlled-$D(\alpha)$ is
\begin{equation}
\label{eq:state}
\frac{1}{\sqrt2}\!\left(\ket0\ket0+iD(\alpha)\ket{-\alpha e^{-i\phi}}\ket1\right)\;.
\end{equation}
It is easy to work out that
\begin{equation}
D(\alpha)\ket{-\alpha e^{-i\phi}}=
D(\alpha)e^{-ia^\dagger a\phi}D(-\alpha)\ket0=
e^{-i|\alpha|^2\sin\phi}\ket{\alpha(1-e^{-i\phi})}\;,
\label{eq:Dstate}
\end{equation}
from which one finds that
\begin{equation}
\matrixel{0}{D(\alpha)}{-\alpha e^{-i\phi}}=
\expect{\alpha}{e^{-ia^\dagger a\phi}}=
e^{-|\alpha|^2(1-e^{-i\phi})}=
e^{-i|\alpha|^2\sin\phi}e^{-2|\alpha|^2\sin^2(\phi/2)}\;.
\end{equation}

In the small-angle approximation,
$|\alpha|^2\phi^2\ll|\alpha|^2\phi\ll1$, the state~(\ref{eq:Dstate})
has negligible overlap with any Fock state other than the vacuum
state and thus is given approximately by $D(\alpha)\ket{-\alpha
e^{-i\phi}}\simeq e^{-i|\alpha|^2\phi}\ket0$.  Written in a slightly
different way, this becomes
\begin{equation}
\ket{\alpha e^{-i\phi}}=e^{-ia^\dagger a\phi}\ket{\alpha}\simeq
e^{-i|\alpha|^2\phi}\ket{\alpha}\;.
\end{equation}
This is the approximation used in the state tracking of
Fig.~\ref{fig16}.  It says that a small phase shift of a coherent
state simply places an appropriate phase in front of the coherent
state; it expresses the fact that for the Fock states with
appreciable amplitude in a coherent state, the phase shifts differ by
roughly $|\alpha|\phi=|\alpha|^2\phi/|\alpha|$, which for small phase
shifts is negligible compared to the average phase shift
$|\alpha|^2\phi$.

The probability for the $Z$ measurement on the qubit to yield outcome
$z$ is $p_z=\frac{1}{2}(1+\langle Z\rangle)$.  The expectation value
of $Z$ in the final state is the same as that for $X$ in the
state~(\ref{eq:state}) before the final Hadamard.  Thus this
probability is given by
\begin{align}
p_z
&=\frac{1}{2}\left[1-z\mbox{Im}
\!\left(\matrixel{0}{D(\alpha)}{-\alpha e^{-i\phi}}\right)\right]\nonumber\\
&=\frac{1}{2}\left(1+z\sin(|\alpha|^2\sin\phi)
e^{-2|\alpha|^2\sin^2(\phi/2)}\right)\nonumber\\
&\simeq\frac{1}{2}\left(1+z\sin(|\alpha|^2\phi)\right)\;.
\label{eq:pzcoh}
\end{align}
The final expression holds in the small-angle approximation and leads
to Heisenberg-limited sensitivity for small phase shifts.

A controlled displacement, like those in Fig.~\ref{fig16}, was
introduced as a potential operation by Davidovich and
co-workers~\cite{Davidovich1993a,Davidovich1996a}, who called it a
{\em quantum switch}, because a qubit coherently switches a classical
driving field on a mode.  These papers suggested that a quantum
switch might be realized at microwave frequencies in a qubit/cavity
set-up like that in Serge Haroche's Paris
laboratory~\cite{Raimond2001a,Davidovich2004a}: a rubidium atom
occupying one of two Rydberg levels, which constitute the qubit,
passes through a superconducting microwave cavity; one level does not
interact with the microwave field, but the other dispersively
switches a cavity mode into resonance with a classical driving field,
which excites the mode into a coherent state.

\begin{figure*}
\center
\includegraphics{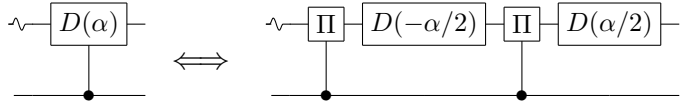}
\caption{A quantum switch, in which a qubit coherently switches a
classical driving field, is equivalent to a sequence of displacements
and controlled parities, as one easily verifies using $D(\alpha/2)\Pi
D(-\alpha/2)=D(\alpha)\Pi$.  A controlled-$\Pi$ is a particular case
of controlled phase shift, with the phase shift set to $\pi$.  The
first controlled-$\Pi$ can be omitted if the input state of the mode
has even parity (e.g., the vacuum state). In our circuit diagrams, we
often use the quantum switch because it is more compact, but the
equivalent procedure is more realistic, at least at microwave
frequencies, where the necessary controlled operations have been
implemented in the laboratory. \label{fig17}}
\end{figure*}

Although a quantum switch has not been implemented in the laboratory,
the same effect can be achieved by using a closely related procedure,
depicted in Fig.~\ref{fig17}, which employs a sequence of
displacements and controlled parities.  A controlled-$\Pi$ is a
particular case of a controlled phase shift, with the phase shift set
to $\pi$.  Such controlled phase shifts have been implemented in the
atom/cavity setting, with the dispersive coupling of one Rydberg
level simply phase shifting the cavity mode.  Such controlled phase
shifts were proposed and analyzed in the atom/cavity setting in
Ref.~\cite{Brune1992a}.  Initial
experiments~\cite{Raimond2001a,Brune1996a} had small phase shifts,
but recent experimental work in Haroche's group takes advantage of
large controlled phase shifts, $\pi/4$~\cite{Guerlin2007a,Brune2008a}
and $\pi$~\cite{Gleyzes2007a,Deleglise2008a}, produced in times much
shorter than the cavity's damping time.  The phase-estimation
procedure of Fig.~\ref{fig16}, with the quantum switch replaced by
the equivalent procedure of Fig.~\ref{fig17}, has been proposed and
analyzed for implementation in the atom/cavity setting by Toscano and
co-authors~\cite{Toscano2006a}.

\begin{figure*}
\center
\includegraphics[width=6.4in]{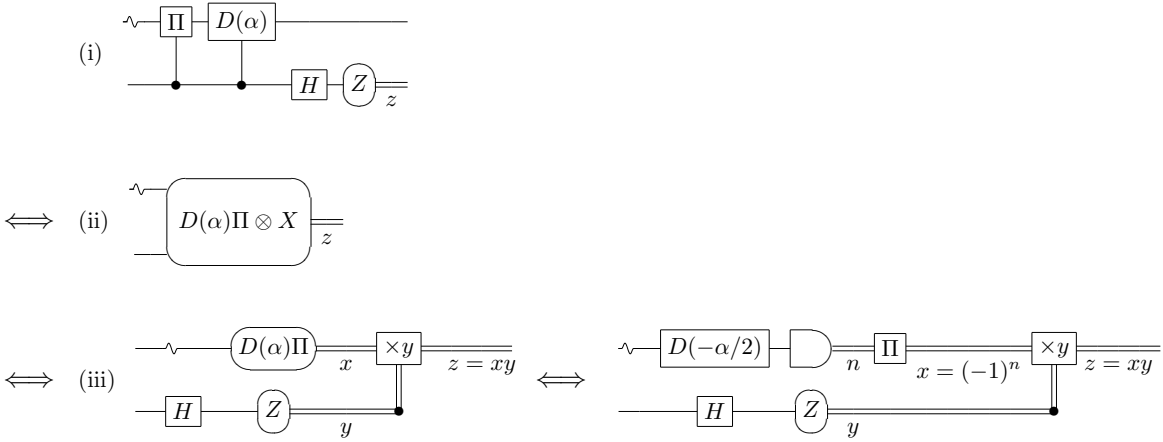}
\caption{Circuit equivalences for disentangling circuits when the
unitary $U$ of Fig.~\protect\ref{fig6}(b) is
$D(\alpha)\Pi=D(\alpha/2)\Pi D(-\alpha/2)$. The controlled-$\Pi$ is
included to make $U$ Hermitian as well as unitary.  Circuits are
numbered as in Fig.~\protect\ref{fig6}(b), but circuit~(iv) is
omitted because it is more productive to guess a coherent preparation
procedure to go with Fig.~\protect\ref{fig16} than to determine an
operator~$P$ that conjugates $D(\alpha)\Pi$ to its opposite.
Circuit~(iii) is converted to a more useful form by replacing the
measurement of $D(\alpha)\Pi$ by a displacement followed by
photocounting and classical extraction of parity. \label{fig18}}
\end{figure*}

\begin{figure*}
\center
\includegraphics[width=6.4in]{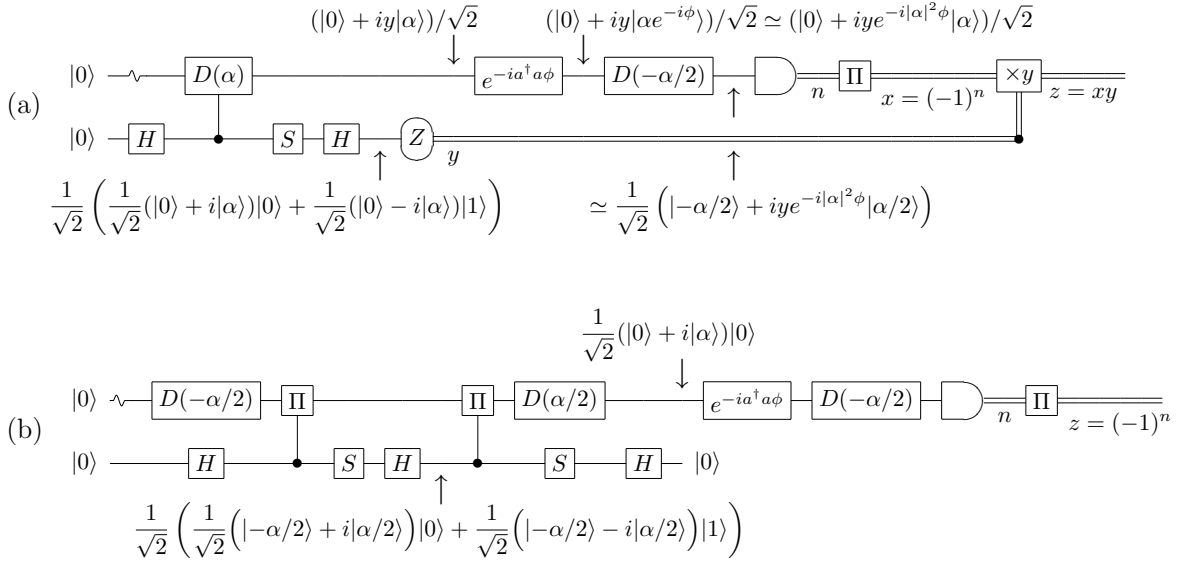}
\caption{(a)~Circuit obtained by applying the second form of
circuit~(iii) in Fig.~\protect\ref{fig18} to the circuit of
Fig.~\protect\ref{fig16}. The controlled-$\Pi$ at the beginning of
Fig.~\protect\ref{fig18}(i) is moved to the beginning of
Fig.~\protect\ref{fig18}(iii).  It is pushed through the phase
shifter, $S$ gate, and controlled-$D(-\alpha)$, converting the last
to a controlled-$D(\alpha)$; it then can be omitted on the mode's
vacuum input.  The superposition state input to the phase shifter is
prepared by post-selection based on the outcome~$y$ of the $Z$
measurement on the qubit.  The two possible outcomes occur with equal
probability; the two input states, $\ket{\psi_{\pm}}=(\ket0\pm
i\ket{\alpha})/\sqrt2$, are normalized, but not orthogonal modal cat
states.  The final state of the mode before photocounting is given in
the small-angle approximation; exact expressions appear in the text.
For small phase shifts, the conditional probability for parity
outcome~$x$ is
$p_{x|y}=\frac{1}{2}(1+x\langle\Pi\rangle_y)\simeq\frac{1}{2}[1+xe^{-|\alpha|^2/2}+xy\sin(|\alpha|^2\phi)]$.
Averaging over $y$ destroys the fringe pattern, but the classical
control circuitry restores a fringe pattern for $z=xy$, with
probability $p_z\simeq\frac{1}{2}(1+z\sin(|\alpha|^2\phi)]$, which
gives Heisenberg-limited sensitivity $\delta\phi=1/|\alpha|$.
(b)~Circuit that coherently prepares modal cat state~$\ket{\psi_+}$
as input to the phase shifter.  Since $D(\alpha/2)\Pi D(-\alpha/2)$
swaps $\ket0$ and $\ket\alpha$, we have $D(\alpha/2)\Pi D(-\alpha/2)
\ket{\psi_\pm}=\pm i\ket{\psi_\mp}$.  Thus we can disentangle the
mode and qubit and prepare $\ket{\psi_+}$ by inserting a
controlled-$\Pi$, preceded by $D(-\alpha/2)$ and succeeded by
$D(\alpha/2)$, just before the qubit measurement in~(a), which can
then be omitted.  Replacing the quantum switch by its equivalent from
Fig.~\protect\ref{fig17} and adding terminal $S$ and Hadamard gates
to the qubit, thereby returning the qubit to its initial state
$\ket0$, yields the circuit shown.  We did not use circuit
equivalences to obtain this coherent circuit, so it is not precisely
equivalent to Fig.~\ref{fig16} and to~(a).  For small phase shifts,
the probability for outcome $z$ is
$p_z\simeq\frac{1}{2}[1+ze^{-|\alpha|^2/2}+z\sin(|\alpha|^2\phi)]$,
which reduces to the outcome probability in Fig.~\ref{fig16} and
in~(a) when $|\alpha|$ is large. \label{fig19}}
\end{figure*}

Our next task is to convert the qubit-assisted protocol of
Fig.~\ref{fig16} to forms that have procedures for making modal cat
states as input to the phase shifter.  In this endeavor,
Fig.~\ref{fig18} specializes the first three disentangling
equivalences of Fig.~\ref{fig6}(b) to the protocol of
Fig.~\ref{fig16}, and Fig.~\ref{fig19} gives protocols in which the
input state to the phase shifter is a modal cat state.

The circuit in Fig.~\ref{fig19}(a) uses post-selection: it prepares
one of two modal cat states,
$\ket{\psi_y}=(\ket0+iy\ket{\alpha})/\sqrt2$, depending on which of
the two equally likely outcomes~$y$ eventuates from the $Z$
measurement on the qubit.  The cat states are the relative states,
$\ket{\psi_0}$ and $\ket{\psi_1}$, of the extended discussion in
Sec.~\ref{subsec:measurements}; they are normalized, but not
orthogonal, having a purely imaginary inner product
$\innerprod{\psi_-}{\psi_+}=i\innerprod{0}{\alpha}=ie^{-|\alpha|^2/2}$.

The protocol of Fig.~\ref{fig19}(a) is precisely equivalent to that
of Fig.~\ref{fig16}, since it is derived by using the disentangling
equivalences of Fig.~\ref{fig18} to disentangle the qubit from the
mode in Fig.~\ref{fig16}.  The state just before the photocounter,
\begin{equation}
\ket{\psi'_y}=
\frac{1}{\sqrt2}D(-\alpha/2)\!\left(\ket0+iy\ket{\alpha
e^{-i\phi}}\right)\;,
\end{equation}
can be used to calculate the probability for outcome $x$ of the
parity measurement, $p_{x|y}=\frac{1}{2}(1+x\expect{\psi'_y}{\Pi})$.
The expectation value of parity is
\begin{equation}
\expect{\psi'_y}{\Pi}=
\frac{1}{2}\!\left(
\innerprod{0}{\alpha}+\matrixel{\alpha e^{-i\phi}}{D(\alpha)}{-\alpha e^{-i\phi}}\right)
-y\mbox{Im}\!\left(\matrixel{0}{D(\alpha)}{-\alpha e^{-i\phi}}\right)\;.
\end{equation}
We need
\begin{equation}
\matrixel{\alpha e^{-i\phi}}{D(\alpha)}{-\alpha e^{-i\phi}}=
e^{-i|\alpha|^2\sin\phi}\innerprod{\alpha e^{-i\phi}}{\alpha(1-e^{-i\phi})}
=e^{-|\alpha|^2/2}e^{-4|\alpha|^2\sin^2(\phi/2)}\;.
\end{equation}
Putting all this together, we get a conditional probability
\begin{align}
\label{eq:pxycoh}
p_{x|y}&=\frac{1}{2}
\left(
1+\frac{1}{2}xe^{-|\alpha|^2/2}(1+e^{-4|\alpha|^2\sin^2(\phi/2)})
+xy\sin(|\alpha|^2\sin\phi)e^{-2|\alpha|^2\sin^2(\phi/2)}
\right)\nonumber\\
&\simeq\frac{1}{2}\!\left(1+xe^{-|\alpha|^2/2}+xy\sin(|\alpha|^2\phi)\right)\;,
\end{align}
where the second expression assumes small phase shifts.  As in the
general discussion in Sec.~\ref{subsec:measurements}, averaging over
$y$ yields an unconditioned probability for $x$ that has no
$\phi$-dependence, but the product variable, $z=xy$, reveals a fringe
pattern, with the unconditioned probability $p_z$ given by
Eq.~(\ref{eq:pzcoh}).

The protocol of Fig.~\ref{fig19}(b) uses a coherent procedure to
prepare the cat state $\ket{\psi_+}$.  It is easy to see that the
probability for outcome~$z$ is given by Eq.~(\ref{eq:pxycoh}) with
$y=+1$ and $x=z$.  This probability is not the same as
Eq.~(\ref{eq:pzcoh}), signaling that the circuit of
Fig.~\ref{fig19}(b) is not quite equivalent to Fig.~\ref{fig16},
though it becomes effectively so when $|\alpha|$ is large, so that
$\ket0$ and $\ket\alpha$ are nearly orthogonal.

Both circuits in Fig.~\ref{fig19} prepare modal cat states for input
to the phase shifter, (a)~by post-selection and (b)~coherently. There
is an independent coherent method for preparing a modal cat state,
which does not require assistance from a qubit, relying instead on a
self-Kerr coupling.   This method, which goes back to the pioneering
work of Yurke and Stoler~\cite{Yurke1986b}, follows mathematically
from the simple identity $e^{-in^2\pi/2}=e^{-i\pi/4}e^{i(-1)^n\pi/4}
=e^{-i\pi/4}[1+i(-1)^n]\sqrt2$, which implies that a self-Kerr
coupling, with a $\pi/2$ phase shift, makes a modal cat state:
\begin{equation}
e^{-i(a^\dagger a)^2\pi/2}\ket{\alpha}=
e^{-|\alpha|^2}\sum_{n=0}^\infty\frac{\alpha^n e^{-in^2\pi/2}}{\sqrt{n!}}\ket n
=\frac{e^{-i\pi/4}}{\sqrt2}(\ket{\alpha}+i\ket{-\alpha})\;.
\end{equation}
The use of Kerr and higher-order nonlinearities to make cat states
and more complicated superpositions of coherent states has been
elaborated extensively~\cite{Tanas1990a,Miranowicz1990a}.

Methods for preparing modal cat states by post-selection have been
proposed using quantum
switches~\cite{Davidovich1993a,Davidovich1996a} and using
controlled phase shifts and controlled nonlinear
interactions~\cite{Gerry1999a,Gerry2000a,Gerry2007b}, and these have
been applied to the cat-state interferometer~\cite{Gerry2000a} of
Fig.~\ref{fig19}(a).  Dalvit and co-workers~\cite{Dalvit2006a} have
investigated post-selected preparation of more elaborate
superpositions of coherent states, such as superpositions of four
coherent states, called compass states~\cite{Zurek2001a}, for use in
interferometers of this sort.  The superpositions are created using
quantum switches and Kerr interactions; the emphasis is on
preparation procedures that might be implemented in an ion trap,
where the qubit is two internal levels of an ion and the mode is a
vibrational mode.  In earlier work, Schneider and
co-workers~\cite{Schneider1998a} reported methods for making such
superpositions of coherent states in an ion trap.

\subsection{Two-mode coherent-state-superposition interferometers}
\label{subsec:twocss}

\begin{figure*}
\center
\includegraphics[width=6.4in]{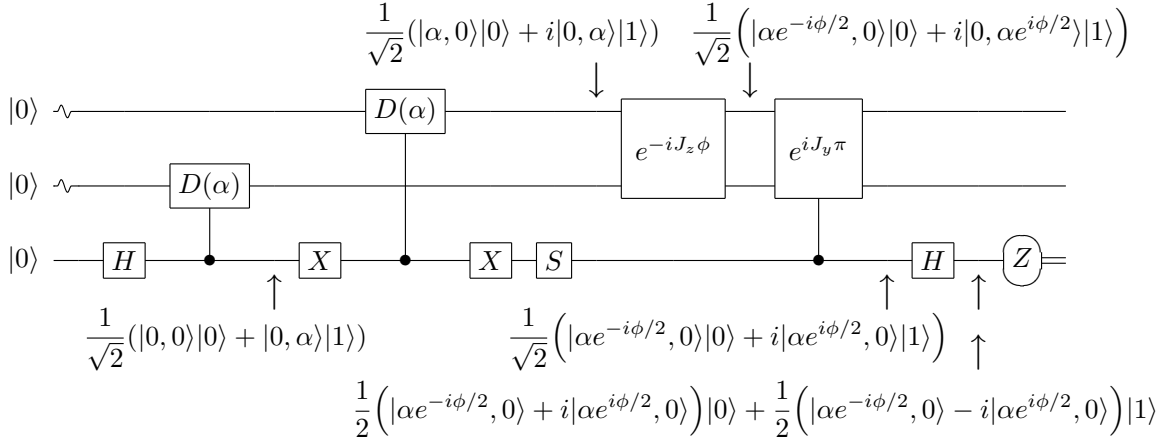}
\caption{Coherent-state analogue of two-mode, qubit-assisted
interferometer in Fig.~\protect\ref{fig10}(d).  To create the
entangled state input to the differential phase shifter, the circuit
uses two quantum switches, the first controlled on $\ket1$ and the
second, via the surrounding $X$ gates, effectively controlled on
$\ket0$.  An $S$ gate on the qubit shifts the fringe pattern by
$\pi/2$ so that optimal sensitivity occurs at $\phi=0$.  The
entangled state input to the beamsplitter is a superposition of the
states at the beamsplitter in Fig.~\protect\ref{fig10}(d), with
coherent-state amplitudes in the superposition and with the extra $i$
to shift the fringe pattern.  Since the remainder of the circuit
preserves photon number, the outcome probability for $z$ can be
obtained by averaging the probability in Fig.~\protect\ref{fig10}(d)
(with a $\pi/2$ phase shift), $\frac{1}{2}(1-z\sin N\phi)$, over the
coherent-state Poisson distribution for $N$; the result is the
distribution~(\protect\ref{eq:pzcoh2}), which gives
Heisenberg-limited sensitivity for small phase shifts. \label{fig20}}
\end{figure*}

Having dealt with single-mode phase estimation, we turn now to the
two-mode, qubit-assisted protocol of Fig.~\ref{fig10}(d).
Figure~\ref{fig20} translates that Fock-state-superposition protocol
into a corresponding coherent-state-superposition protocol.  The
translation replaces the preparation procedure of Fig.~\ref{fig10}(d)
with a procedure that uses two quantum switches to make the
appropriate mode-qubit entangled state, as proposed in
Refs.~\cite{Davidovich1993a,Davidovich1996a}.  The rest of the
circuit remains the same, except for the addition of an $S$ gate on
the qubit wire, which shifts the fringe pattern by $\pi/2$ so that
$\phi=0$ is an optimal operating point.  This turns out to be a
natural addition, just as in Sec.~\ref{subsec:onecss}. The final
state provides the probability for outcome~$z$,
\begin{align}
p_z&=\frac{1}{4}
\left|\left|\ket{\alpha e^{-i\phi/2}}+iz\ket{\alpha e^{i\phi/2}}\right|\right|^2\nonumber\\
&=\frac{1}{2}\left[1+z\mbox{Im}\!\left(\matrixel{\alpha}{e^{-ia^\dagger a\phi}}{\alpha}\right)\right]\nonumber\\
&=\frac{1}{2}\left(1-z\sin(|\alpha|^2\sin\phi)e^{-2|\alpha|^2\sin^2(\phi/2)}\right)\;,
\label{eq:pzcoh2}
\end{align}
which gives Heisenberg-limited sensitivity for small phase shifts.

\begin{figure*}
\center
\includegraphics[width=6.75in]{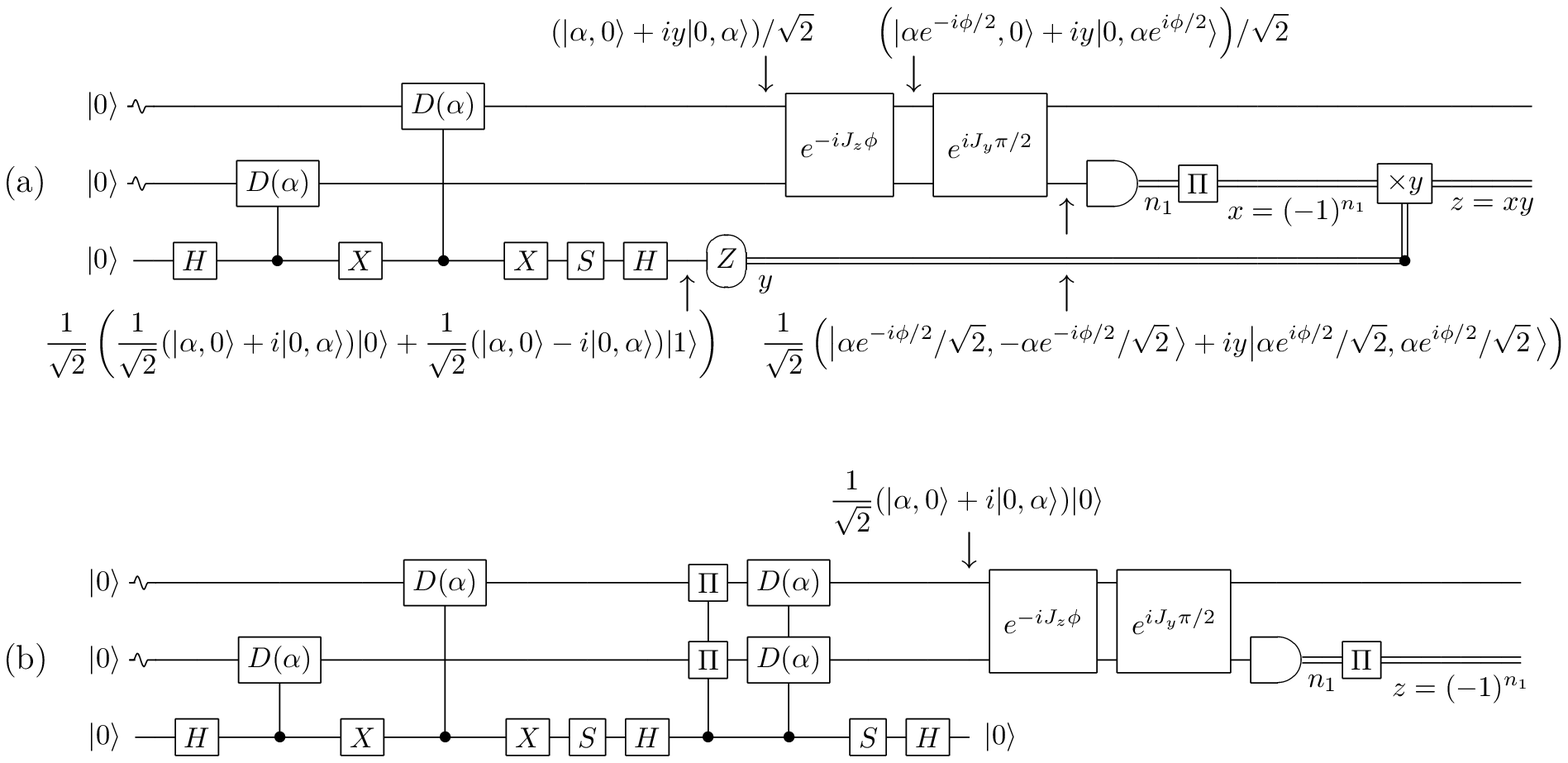}
\caption{(a)~Circuit obtained by applying the second form of
circuit~(iii) in Fig.~\protect\ref{fig14} to the circuit of
Fig.~\protect\ref{fig20}. The controlled-$\Pi$ at the beginning of
Fig.~\protect\ref{fig14}(i) is shifted to the beginning of
Fig.~\protect\ref{fig14}(iii).  It can be pushed through to just
after the initial Hadamard on the qubit wire and can then be omitted
on mode~0's vacuum input.  The superposition state input to the phase
shifter is prepared by post-selection based on the outcome~$y$ of the
$Z$ measurement on the qubit.  The two equally likely outcomes
produce input 0BB0 states, $\ket{\Psi_{\pm}}=(\ket{\alpha,0}\pm
i\ket{0,\alpha})/\sqrt2$, which are normalized, but not orthogonal;
they are superpositions, with coherent-state amplitudes, of N00N
states, but with an additional $i$ phase shift.  Since the remainder
of the circuit preserves photon number, we can calculate the
probability for outcome~$z$ as a Poisson average of the N00N state
probability of Fig.~\protect\ref{fig15}(a) (with an additional
$\pi/2$ phase shift), $p_z=\frac{1}{2}(1-z\sin N\phi)$; the result is
the probability $p_z$ of Eq.~(\protect\ref{eq:pzcoh2}).  (b)~Circuit
that coherently prepares 0BB0 state~$\ket{\Psi_+}$ as input to the
differential phase shifter. Since $D(\alpha)\Pi$ swaps $\ket0$ and
$\ket\alpha$, we can disentangle the mode and qubit and prepare
$\ket{\Psi_+}$ by inserting on both modes a controlled-$\Pi$ followed
by a controlled-$D(\alpha)$.  Adding terminal $S$ and Hadamard gates
to the qubit, thereby returning it to its initial state $\ket0$,
gives the circuit shown.  Since we did not use circuit equivalences
to obtain this coherent circuit, it is not precisely equivalent to
Fig.~\protect\ref{fig20} and to~(a), though it becomes so when
$|\alpha|$ is large.
 \label{fig21}}
\end{figure*}

The next task is to convert the qubit-assisted protocol of
Fig.~\ref{fig20} to forms that have procedures for making the
appropriate mode-entangled states, $(\ket{\alpha,0}\pm
i\ket{0,\alpha})/\sqrt2$, as input to the differential phase shifter.
The resulting circuits are given in Fig.~\ref{fig21}. In CMC's
research group at the University of New Mexico, we have taken to
referring to these states as 0BB0 states (pronounced ``oboe''),
because we often write them (unnormalized) as
$\ket{0,\beta}+\ket{\beta,0}$~\cite{0BB0}.  We have analyzed the
performance of 0BB0 states under losses and considered schemes for
implementation, all of which is to be presented
elsewhere~\cite{Datta2009au}.

The circuit in Fig.~\ref{fig21}(a) uses post-selection: it prepares
one of the two 0BB0 states,
$\ket{\Psi_y}=(\ket{\alpha,0}+iy\ket{0,\alpha})/\sqrt2$, depending on
which of the two (equally likely) outcomes $y$ emerges from the $Z$
measurement on the qubit.  The two 0BB0 states are normalized, but
they are not orthogonal, their inner product,
$\innerprod{\Psi_-}{\Psi_+}=i|\innerprod{0}{\alpha}|^2=ie^{-|\alpha|^2}$,
being purely imaginary.

\begin{figure*}
\center
\includegraphics{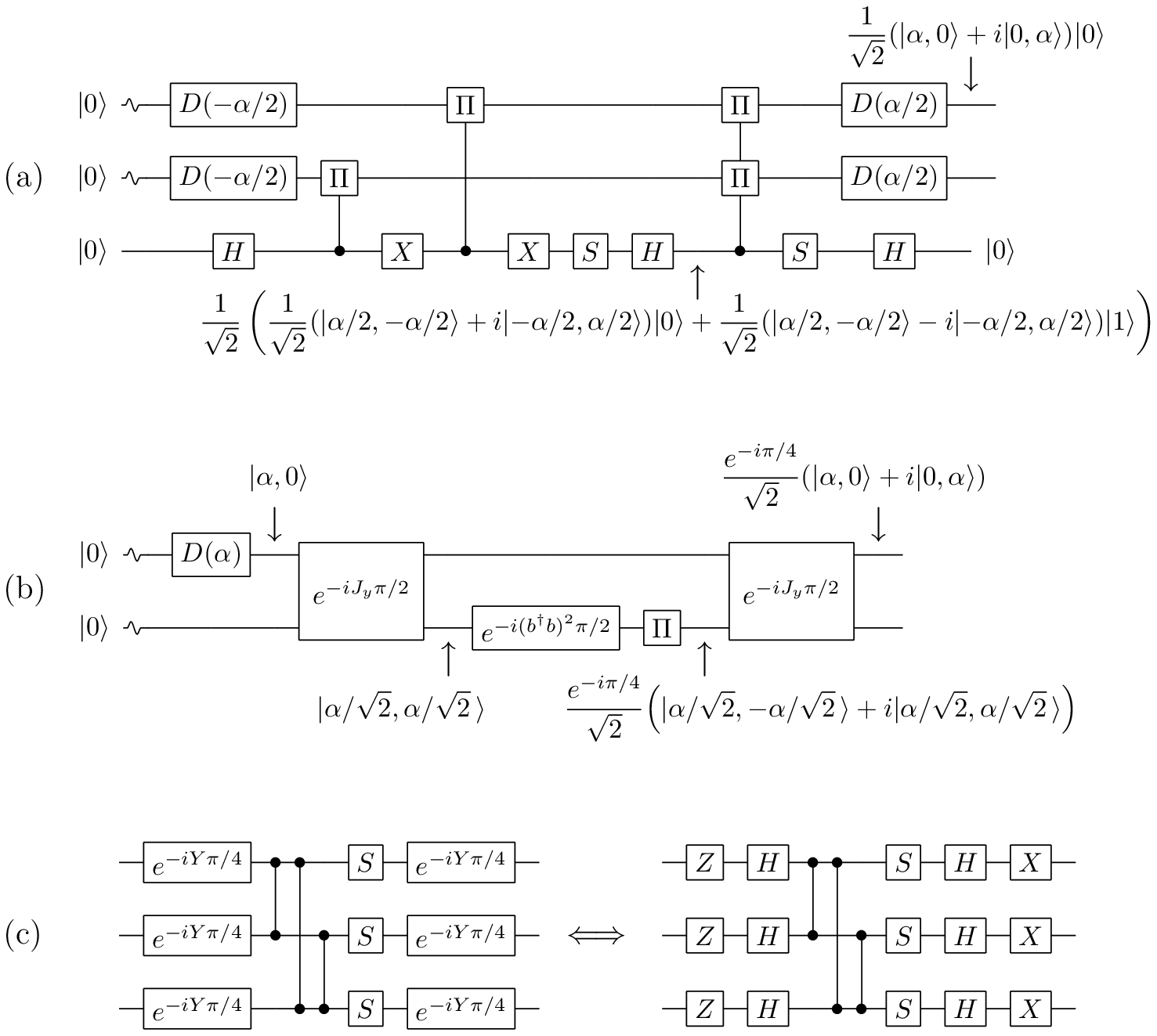}
\caption{Coherent preparation procedures for 0BB0 states. (a)~The
preparation part of the circuit in Fig.~\ref{fig21}(b), with quantum
switches eliminated in favor of controlled parities by using the
equivalence of Fig.~\protect\ref{fig17}. (b)~Kerr-gate protocol for
making 0BB0 states.  The circuit is a nonlinear interferometer, with
50/50 beamsplitters before and after a Kerr self-coupling and a $\pi$
phase shift on mode~1. (c)~Translation to $N=3$ qubits of the part of
circuit~(b) after the displacement of mode~0.  The translation uses
the circuit equivalence of Fig.~\protect\ref{fig2}(a); the
equivalence uses $e^{-iY\pi/4}=HZ=XH$.  Since the part of~(b) after
the displacement of mode~0 preserves photon number, if
$\ket{\alpha,0}$ is replaced by $\ket{N,0}$ at this point in the
circuit, the output is the N00N-like state
$e^{-i\pi/4}(\ket{N,0}+i\ket{0,N}/\sqrt2$.  Equivalently, (c) takes
input $\ket0^{\otimes N}$ to the cat-like
state~$e^{-i\pi/4}(\ket0^{\otimes N}+i\ket1^{\otimes N})/\sqrt2$.  On
this input, the initial $Z$s can be omitted, and the final $X$s only
produce an overall phase, so they, too, can be omitted.
\label{fig22}}
\end{figure*}

The circuit of Fig.~\ref{fig21}(a) is precisely equivalent to that of
Fig.~\ref{fig20}, since it can be derived from Fig.~\ref{fig20} using
the disentangling equivalences of Fig.~\ref{fig14}.  The state just
before the photocounter,
\begin{equation}
\ket{\Psi'_y}=
\frac{1}{\sqrt2}\left(
\bket{\alpha e^{-i\phi/2}/\sqrt2,-\alpha e^{-i\phi/2}/\sqrt2}
+iy\bket{\alpha e^{i\phi/2}/\sqrt2,\alpha e^{i\phi/2}/\sqrt2}
\right)\;,
\end{equation}
provides the probability for outcome $x$ of the parity measurement,
$p_{x|y}=\frac{1}{2}(1+x\expect{\Psi'_y}{I\otimes\Pi})$, via the
expectation value of the parity of mode~1,
\begin{align}
\expect{\Psi'_y}{I\otimes\Pi}&=
\innerprod{0}{\sqrt2\alpha}+\mbox{Im}\!\left(
\matrixel{\alpha/\sqrt2}{e^{-ia^\dagger a\phi}}{\alpha/\sqrt2\,}^2
\right)\nonumber\\
&=e^{-|\alpha|^2}-y\sin(|\alpha|^2\sin\phi)e^{-2|\alpha|^2\sin^2(\phi/2)}\;.
\end{align}
Thus the conditional probability is
\begin{equation}
\label{eq:pxycoh2}
p_{x|y}=\frac{1}{2}
\left(
1+xe^{-|\alpha|^2}
-xy\sin(|\alpha|^2\sin\phi)e^{-2|\alpha|^2\sin^2(\phi/2)}
\right)
\simeq\frac{1}{2}\!\left(1+xe^{-|\alpha|^2}-xy\sin(|\alpha|^2\phi)\right)\;,
\end{equation}
where the second expression is valid for small phase shifts. Although
the unconditioned probability for $x$ does not depend on $\phi$, a
fringe pattern appears in the product variable, $z=xy$, whose
unconditioned probability is given by Eq.~(\ref{eq:pzcoh2}).

The corresponding protocol that prepares 0BB0 states coherently is
depicted in Fig.~\ref{fig21}(b).  The probability for outcome~$z$ is
given by the conditional probability~(\ref{eq:pxycoh2}) with $y=+1$
and $x=z$.  This probability is not the same as the final outcome
probability in Fig.~\ref{fig20}, telling us that this coherent
protocol is not precisely equivalent to the circuits of
Fig.~\ref{fig20} and Fig.~\ref{fig21}(a).  It becomes equivalent when
$|\alpha|$ is large, so that $\ket{\alpha,0}$ and $\ket{0,\alpha}$
are essentially orthogonal.

The preparation procedure of Fig.~\ref{fig21}(b), which employs
quantum switches, is converted in Fig.~\ref{fig22}(a) to an
equivalent protocol that uses controlled parities.   This protocol is
perhaps a practical way to make an 0BB0 state at microwave
frequencies because of the ability in an atom/cavity setting to do
displacements and controlled phase shifts in times much shorter
than the time scale for cavity damping.

Figure~\ref{fig22}(b) provides another, entirely independent coherent
preparation procedure for 0BB0 states, which was proposed by Gerry,
Benmoussa, and Campos~\cite{Gerry2002a}.  This procedure is a
nonlinear interferometer, with a self-Kerr coupling in one arm.  It
generates an 0BB0 state by first making a modal cat state with the
self-Kerr coupling and using the 50/50 beamsplitters of the
interferometer to turn the cat state into an 0BB0 state.  The
nonlinear interferometer preserves photon number and thus can be
converted to qubits using the equivalence of Fig.~\ref{fig2}(a); the
resulting qubit interferometer, which relies on controlled-SIGN gates
between all pairs of qubits, is depicted in Fig.~\ref{fig22}(c).  The
equivalent circuits~(b) and~(c) have been suggested and analyzed by
Gerry and Campos~\cite{Gerry2003a} as a method for making a N00N-like
state in a~\hbox{BEC}.

\section{Conclusion}
\label{sec:conclusion}

After all this discussion of protocols for phase estimation, what can
we say about the prospects for implementing Heisenberg-limited phase
estimation in the lab?  Probably the best place to try is at
microwave frequencies in a qubit/cavity setting, where coherent
states can be manipulated relatively easily and where appropriate
controlled operations are already available.  One might think about
performing one of the two-mode coherent-state protocols of
Sec.~\ref{subsec:twocss}.  The qubit-assisted protocol of
Fig.~\ref{fig20} requires a controlled swap of the two modes, and the
0BB0-preparing protocols of Figs.~\ref{fig21} and \ref{fig22}(a)
require 50/50 beamsplitters and a measurement of parity.  These are
tall orders, except the parity measurement, which can be done, for
example, by mapping to an ancillary qubit as in Fig.~\ref{fig6}(c).

If one's primary goal, however, is Heisenberg-limited phase
estimation---and not making 0BB0 states---there is little reason to
use two microwave modes.  Almost certainly the most practical
protocols are the single-mode coherent-state-superposition protocols
of Sec.~\ref{subsec:onecss}.  One can imagine implementing the
qubit-assisted, single-mode phase-estimation circuit of
Fig.~\ref{fig16}, with the quantum switches replaced by
controlled parities, or perhaps the corresponding modal-cat-state
protocols of Fig.~\ref{fig19}, with the quantum switches again
removed in favor of controlled parities and the final parity
measurement performed using an ancillary qubit.

Despite the title of this paper, what we deal with here is not so
much interferometry as general phase estimation in interferometric
and other settings.  Generally, we think about the phase shift as
being applied to atoms, as in Ramsey interferometry, or to modes of
the electromagnetic field, as in optical interferometry.

In a conventional, quantum-noise-limited interferometer, there is a
$\pi/2$ pulse or beamsplitter immediately before and immediately
after the application of the phase shift.  Before the initial
beamsplitter, there is some relatively simple state preparation, and
after the final beamsplitter, there is a relatively simple
measurement.  Once we generalize to Heisenberg-limited phase
estimation, however, we outgrow these beamsplitters.  We must
consider any preparation procedure before the phase shifter and any
measurement procedure after it.  In this more general setting, it is
clearly better to use a representation that is not tied to the two
beamsplitters and that can represent whatever goes on before and
after the phase shifter.

This is the reason that quantum circuits provide a far more useful
way of representing general protocols for phase estimation and
interferometry.  Quantum circuits provide a platform-independent
pictorial representation of all phase-estimation protocols, allowing
us to switch easily from qubits to modes and to visualize---almost to
experience, so tactile are the circuit diagrams---the equivalences
between different sorts of interferometers and other phase-estimation
protocols.  The goal of this paper has been to develop tools for
manipulating the quantum circuits for phase estimation efficiently.

\acknowledgements

This article is dedicated to the memory of Krzysztof W{\'o}dkiewicz,
friend, colleague, and New Mexican.  During the time CMC has been at
the University of New Mexico, since 1992, Krzysztof spent every other
year in Albuquerque.  He was an exemplar of excellence, taste, rigor,
style, and vitality in doing theoretical physics, both in his
teaching and in his research.  CMC~will always remember him by the way
he reappeared in August on his return to UNM after a year in Warsaw:
he would bounce down the hallway and burst into the office, eager to
share his latest research, always with the same infectious
enthusiasm.  He approached his work in physics with a vitality that
never faltered, and {\em that\/} is perhaps the best of several
reasons that we will always remember him as a model of a theoretical
physicist.

CMC thanks the School of Physical Sciences of the University
of Queensland for hospitality when this work was getting started.
This work was supported in part by the U.S.~Office of Naval Research
(Grant No.~\hbox{N00014-07-1-0304}).  Quantum circuits in this paper
were drawn using a modified version of B.~Eastin and S.~T.~Flammia's
Qcircuit package, which draws circuits in LaTeX; the modified version,
provisionally called Qcircuitw, is available from the authors.


\begin{references}
\bibitem{fJ}
This is because the algebra of the operators $I$, $J_z$, and
$J_{\pm}=J_x\pm iJ_y$ spans the space of operators on the space of
angular momentum $J=N/2$.  One way to see this is to notice that any
outer product in the angular-momentum basis, $\outerprod{J,m}{J,m'}$,
is a multiple of $J_-^{J-m}\proj{J,J}J_+^{J-m'}$, which in turn is a
multiple of $J_-^{J-m}J_+^{2J}J_zJ_-^{2J}J_+^{J-m'}$.
\bibitem{0BB0}
We studiously avoid the N00N-state ordering for 0BB0 states.
\end{references}
\end{document}